\documentclass[prd,twocolumn,showpacs,showkeys,letterpaper,amsmath,amssymb,floatfix]{revtex4}
\usepackage{natbib}
\usepackage{amsmath}
\usepackage{amssymb}

\usepackage{dcolumn}
\usepackage{bm}

\newcommand{\z}[0]{{\text z}}
\newcommand{\ta}[0]{{t_\text{o}}}
\newcommand{\bp}[0]{{\beta_\text{o}}}
\newcommand{\da}[0]{{d_\text{o}}}
\newcommand{\lp}[0]{{\lambda_\text{o}}}
\newcommand{\ie}[0]{{i.e.}, }

\newcommand{\figref}[1]{{figure~\ref{#1}}}
\newcommand{\Figref}[1]{{Figure~\ref{#1}}}
\newcommand{\eqnref}[1]{{equation~\eqref{#1}}}
\newcommand{\Eqnref}[1]{{Equation~\eqref{#1}}}
\newcommand{\eqnsref}[2]{{equations~\eqref{#1} and \eqref{#2}}}
\usepackage{graphics}
\newcommand{\image}[4]
{
\begin{figure}[htb]
\centering
\makebox{
  \resizebox
      {#2cm}{!}
     {\includegraphics{#1}}
}
\caption{#4}
\label{#3}
\end{figure}
}
\def\astrobj#1{#1}

\begin{document}

\title{Constraints of Perception and Cognition in Relativistic
  Physics}

\author{Manoj Thulasidas}

\affiliation{Neural Signal Processing Lab, Institute for Infocomm
  Research,\\
 National University of Singapore,
21 Heng Mui Keng Terrace, Singapore 119613.}

\email{manoj@i2r.a-star.edu.sg}

\date{\today}

\label{firstpage}

\begin{abstract}
  Cognitive neuroscience treats reality as our brain's representation
  of our sensory inputs.  In this view, our perceptual reality is only
  a distant and convenient mapping of the physical processes causing
  the sensory inputs.  Sound is a mapping of auditory inputs, and
  space is a representation of visual inputs. Any limitation in the
  chain of sensing has a specific manifestation on the cognitive
  representation that is our perceived reality.  One physical
  limitation of our visual sensing is the finite speed of light.  The
  manifestation of this limitation is the reason why the speed of
  light appears at the basic structure of our space-time. Physics,
  however, treats the perceptual reality of space and time (our
  brain's representation) as a faithful image of the physical reality
  (the objects and phenomena causing the sensory inputs).  This faith
  in our cognitive map or perceived reality results in attributing the
  manifestations of the limitations of our perception to the true
  nature of space and time.  In this article, we look at the
  consequences of the limited speed of our perception, namely the
  speed of light, and show that they are remarkably similar to the
  coordinate transformation in the special theory of relativity.
  Further illustrating the validity of looking at the visual reality
  as our brain's representation limited by the speed of light, we show
  that we can unify and explain a wide array of seemingly unrelated
  astrophysical and cosmological phenomena using this framework.
  Understanding the constraints on our space and time due to the
  limitations in perception and cognitive representation opens the
  possibility of understanding astrophysics and cosmology from a whole
  new viewpoint.

\end{abstract}



\maketitle

\section{Introduction}\label{S1}

Our reality is a mental picture that our brain creates, starting from
our sensory inputs \citep{vr}. Although this cognitive map is often
mistaken to be a faithful image of the physical causes behind the
sensing process, the causes themselves are very different from the
perceptual experience of sensing.  The difference between the
cognitive representation and their physical causes is not immediately
obvious when we consider our primary and most powerful sense of sight.
But, we can appreciate the difference by looking at our less powerful
olfactory and auditory senses.  Odors, which may appear to be a
property of the air we breathe, are in fact our brain's representation
of the chemical signatures that our nose senses.  Similarly, sound is
not an intrinsic property of a vibrating body, but our brain's
mechanism to represent the pressure waves in the air.
Table~\ref{tab:rep} shows the chain from the physical cause of the
sensory input to the final reality as the brain creates it.  Although
the physical causes can be identified for olfactory and auditory
chains, they are not easily discerned for visual process. Since sight
is the most powerful sense we possess, we are obliged to accept our
brain's representation of visual inputs as the fundamental reality.

\begin{table}[htb]
{
\centering
\begin{tabular}{ c   c   c   c }
\hline
\hline
      Sense & Physical & Sensing & Brain's   \\
      modality & cause & signal & \quad representation \quad  \\
\hline
\quad Olfactory\quad & Chemicals & Chemical & Smells \\
          &          & \quad concentrations \quad & \\
\hline
Auditory & \quad Vibrating  \quad & \quad Air pressure \quad &  Sounds \\
         & objects &      waves & \\
\hline
Visual & Unknown & Light & Space, time \\
         &        &        & reality \\
\hline
\hline
\end{tabular}
}
\caption[Brain's representation of different sensory inputs.]
{ Brain's representation of different sensory inputs.  Odors are a representation of
  chemical compositions and concentration our nose senses. Sounds are
  a mapping of the air pressure waves produced by a vibrating object.
  In sight, we do not know the physical reality, our representation is
  space, and possibly time.\label{tab:rep}}

\end{table}

A good indication of the tight integration between the physiology of
perception and its representation in the brain was proven recently
\citep{mapping} in a clever experiment using the tactile funneling
illusion.  This illusion results in a single tactile sensation at the
focal point at the center of a stimulus pattern even though no
stimulation is applied at that site.  In the experiment, the brain
activation region corresponded to the focal point where the sensation
was perceived rather than the points where the stimuli were applied,
proving that brain registered perceptions, not the physical causes of
the perceived reality.  In other words, for the brain, there is no
difference between applying the pattern of the stimuli and applying
just one stimulus at the center of the pattern.  Brain maps the
sensory inputs to regions that correspond to their perception, rather
than the regions that physiologically correspond to the sensory
stimuli.

The neurological localization of different aspects of reality has been
established by lesion studies in neuroscience.  The perception of
motion (and the consequent basis of our sense of time), for instance,
is so localized that a tiny lesion can erase it completely.  Cases of
patients with such specific loss of a part of reality \cite{vr}
illustrate the fact that our experience of reality, every aspect of
it, is indeed a creation of the brain.  Space and time are aspects of
the cognitive representation in our brain.

Space is a perceptual experience much like sound. Comparisons between
the auditory and visual modes of sensing can be useful in
understanding the limitations of their representations in the brain.
One limitation is the input ranges of the sensory organs.  Ears are
sensitive in the frequency range 20Hz--20kHz, and eyes are limited to
the visible spectrum.  Another limitation, which may exist in specific
individuals, is an inadequate representation of the inputs.  Such a
limitation can lead to tone-deafness and color-blindness, for
instance. The speed of the sense modality also introduces an effect,
such as the time lag between seeing an event and hearing the
corresponding sound.  For visual perception, the consequence of the
finite speed of light is called the light travel time effect, which
offers a convincing explanation to the observed superluminal motion in
certain celestial objects \citep{M87,superluminal2}.  When an object
approaches the observer at a shallow angle, the transverse speed may
appear superluminal.

However, other consequences of the light travel time (LTT) effect are
very similar to the coordinate transformation of the special theory of
relativity (SR).  These consequences include an apparent contraction
of a receding object along its direction of motion and a time dilation
effect.  Furthermore, a receding object can never {\em appear} to be
going faster than the speed of light, even if its real speed is
superluminal.  While SR does not explicitly forbid it, superluminality
is understood to lead to time travel and the consequent violations of
causality.  An {\em apparent} violation of causality is one of the
consequences of LTT, when the superluminal object is approaching the
observer.  All these effects due to LTT are remarkably similar to SR.
However, LTT effects are currently assumed to apply on a space-time
that obeys SR.  It may be that there is a deeper structure to the
space-time, of which SR is only our perception, filtered through
LTT effect.  By treating LTT effects as an optical illusion to be
applied on an SR-like space-time, we may be double counting them.

Once we accept the neuroscience view of reality as a representation of
our sensory inputs, we can understand why the speed of light figures
so prominently in our physical theories.  The theories of physics are
a description of reality.  Reality is created out of the readings from
our senses, especially our eyes. They work at the speed of light.
Thus the sanctity accorded to the speed of light is a feature only of
{\em our} reality, not the absolute, ultimate reality which our senses
are striving to perceive.  When it comes to physics that describes
phenomena well beyond our sensory ranges, we really have to take into
account the role that our perception and cognition play in seeing
them. The universe as we see it is only a cognitive model created out
of the photons falling on our retina or on the photo-sensors of the
Hubble telescope.  Because of the finite speed of the information
carrier (namely photons), our perception is distorted in such a way as
to give us the impression that space and time obey special relativity.
They do, but space and time are not the absolute reality. ``Space and
time are modes by which we think and not conditions in which we
live,'' as Einstein himself put it.

Treating our perceived reality as our brain's representation of our
visual inputs (filtered through the light travel time effect), we will
see that all the strange effects of the coordinate transformation in
special relativity can be understood as the manifestations of the
finite speed of our senses in our space and time.  Furthermore, this
line of thinking leads to natural explanations for two classes of
astrophysical phenomena:
\begin{description}
\item[Gamma Ray Bursts] currently believed to emanate from
  cataclysmic stellar collapses, and
\item[Radio Sources] considered manifestations of space-time
  singularities or neutron stars.
\end{description}
Beyond unifying these apparently distinct astrophysical phenomena, the
cognitive limitations to reality can provide qualitative explanations
for such cosmological features as the apparent expansion of the
universe and the Cosmic Microwave Background Radiation (CMBR).  Both
these phenomena can be understood as related to our perception of
superluminal objects.  It is the unification of these seemingly
distinct phenomena at vastly different length and time scales, along
with its conceptual simplicity, that we hold as the indicators of
validity of this framework.

\section{Similarities between LTT Effects and SR}

The coordinate transformation derived in the original paper
\citep{einstein} is, in part, a manifestation of light travel time
(LTT) effects.  This is most obvious in the first thought experiment,
where observers moving with a rod find their clocks not synchronized
due to the difference in light travel times along the length of the
rod.  In this section, we will consider space and time as a part of
the cognitive model created by the brain, and illustrate that special
relativity applies to the cognitive model.  The absolute reality (of
which the SR-like space-time is our perception) does not have to obey
the restrictions of SR.  In particular, objects are not restricted to
subluminal speeds, but they may appear to us as though they are
restricted to subluminal speeds in our perception of space and time.
If we disentangle LTT effects from the rest of SR, we can
understand a wide array of phenomena, as we shall see in this article.
Although not attempted in this paper, the primary motivation for SR,
namely the covariance of Maxwell's equations, may be accomplished even
without attributing LTT effects to the properties of space and
time.

A feature of SR is that it seeks a linear coordinate transformation
between coordinate systems in motion with respect to each other.  We
can trace the origin of linearity to a hidden assumption on the nature
of space and time built into SR, as stated by Einstein
\citep{einstein}: ``In the first place it is clear that the equations
must be linear on account of the properties of homogeneity which we
attribute to space and time.''  Because of this assumption of
linearity, the original derivation of the transformation equations
ignores the asymmetry between approaching and receding objects.  Both
approaching and receding objects can be described by two coordinate
systems that are always receding from each other.  For instance, if a
system $K$ is moving with respect to another system $k$ along the
positive X axis of $k$, then an object at rest in $K$ at a positive
$x$ is receding while another object at a negative $x$ is approaching
an observer at the origin of $k$.  Unlike SR, considerations based on
LTT effects result in intrinsically different set of
transformation laws for objects approaching an observer and those
receding from him. More generally, the transformation depends on the
angle between the velocity of the object and the observer's line of
sight. Since the transformation equations based on LTT effects
treat approaching and receding objects asymmetrically, they provide a
natural solution to the twin paradox, for instance.

\subsection{Perception of Speed}

We first look at how the perception of motion is modulated by the
light travel time (LTT) effects. As remarked earlier, the
transformation equations of SR treat only objects receding from the
observer.  For this reason, we first consider a receding object,
flying away from the observer at a speed $\beta = v/c$, where $c$ is
the speed of light.  The apparent speed $\bp$ of the object depends on
the real speed $\beta$ (as shown in Appendix~\ref{sec.bp}):
\begin{eqnarray}
\nonumber
\bp \,&=&\, \frac{\beta}{1\,+\,\beta} \\
\nonumber
\lim_{\beta\to\infty} \bp \,&=&\, 1
\end{eqnarray}
Thus, due to LTT effects, an infinite real velocity gets
mapped to an apparent velocity $\bp=1$. In other words, no object can
appear to travel faster than the speed of light, entirely consistent
with SR.

Physically, this apparent speed limit amounts to a mapping of $c$
to $\infty$, which is most obvious in its consequences.  For instance,
it takes an infinite amount of energy to accelerate an object to an
apparent speed $\bp=1$, because, in reality, we are accelerating it to
an infinite speed. This infinite energy requirement can also be viewed
as the relativistic mass changing with speed, reaching $\infty$ at
$\bp=1$.  Einstein explained this mapping as: ``For velocities greater
than that of light our deliberations become meaningless; we shall,
however, find in what follows, that the velocity of light in our
theory plays the part, physically, of an infinitely great velocity.''
Thus, for objects receding from the observer, the effects of LTT are
almost identical to the consequences of SR, in terms of the perception
of speed.

\subsection{Time Dilation}
\image{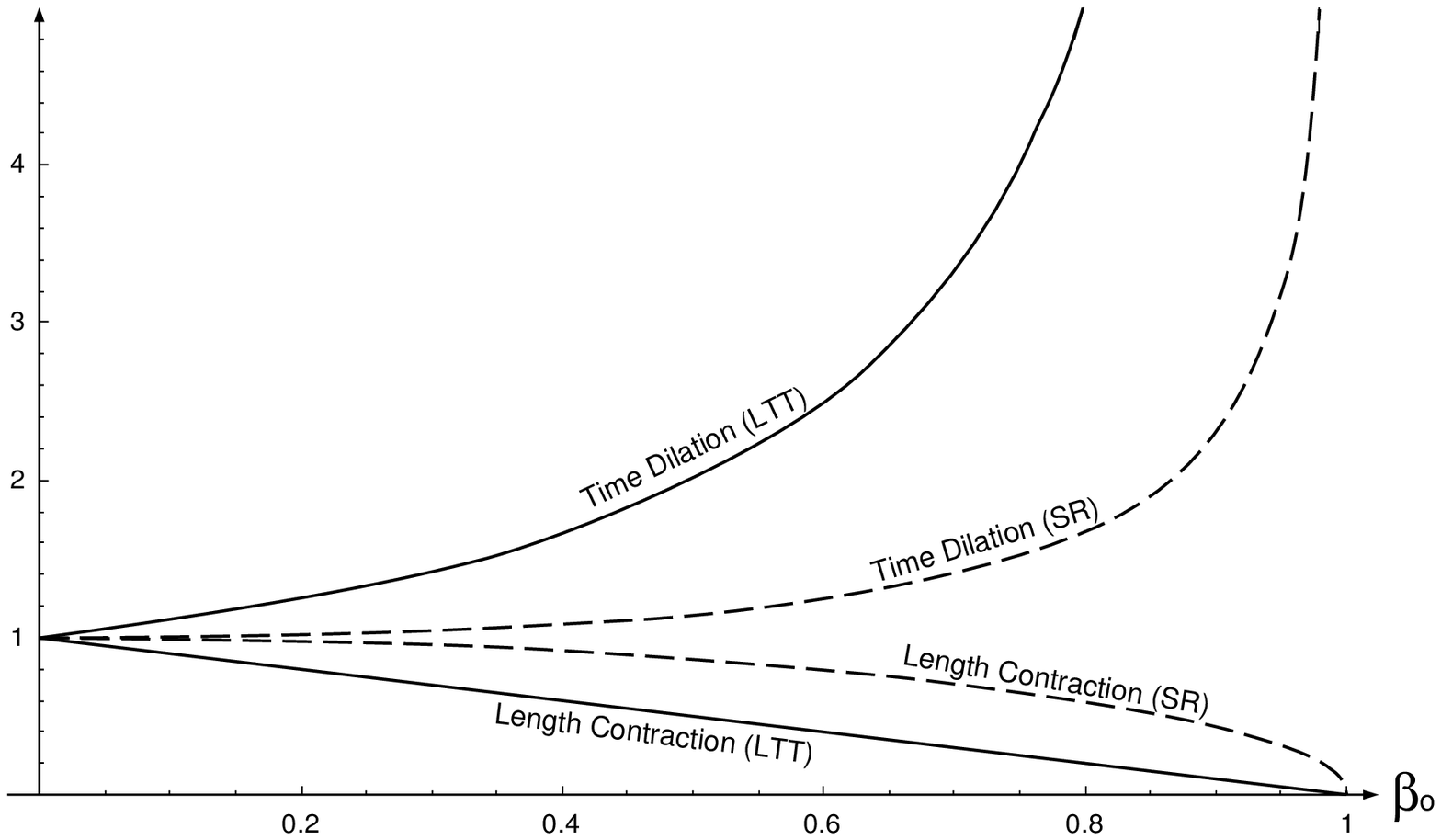}{8}{LTvsSR} {Comparison between light travel time (LTT)
  effects and the predictions of the special theory of relativity
  (SR).  The X-axis is the apparent speed and the Y-axis shows the
  relative time dilation or length contraction.}

LTT effects influence the way time at the moving object is perceived.
Imagine an object receding from the observer at a constant rate.  As
it moves away, the successive photons emitted by the object take
longer and longer to reach the observer, because they are emitted at
farther and farther away.  This travel time delay gives the observer
the illusion that time is flowing slower for the moving object.  It
can be easily shown (see Appendix~\ref{sec.tp}) that the time interval
observed $\Delta\ta$ is related to the real time interval $\Delta t$
as:
\begin{eqnarray}
\nonumber
  \frac{\Delta\ta}{\Delta t} &\,=\,& \frac{1}{1-\bp}
\end{eqnarray}
for an object receding from the observer ($\theta=\pi$).  This
observed time dilation is plotted in \figref{LTvsSR}, where it is
compared to the time dilation predicted in SR.  Note that the time
dilation due to LTT is stronger than the one predicted in SR. However,
the variation is similar, with both time dilations tending to $\infty$
as the observed speed tends to $c$.

\subsection{Length Contraction}
The length of an object in motion also appears different due to LTT
effects.  It can be shown (see Appendix~\ref{sec.lp}) that observed
length $\da$ is related to the real length $d$ as:
\begin{eqnarray}
\nonumber
\frac{\da}{d} &\,=\,& {1-\bp}
\end{eqnarray}
for an object receding from the observer with a speed of $\bp$.  This
equation also is plotted in \figref{LTvsSR}.  Note again that the LTT
effects are stronger than the ones predicted in SR.

\Figref{LTvsSR} illustrates that both time dilation and Lorentz
contraction can be thought of as LTT effects.  While the actual
magnitudes of LTT effects are larger than what SR predicts, their
qualitative behavior as a function of speed is very similar.  This
similarity is not surprising because the coordinate transformation in
SR is partly based on light travel time effects.  If LTT effects are
to be applied, as an optical illusion, on top of the consequences of
SR as currently believed, then the total observed length contraction
and time dilation will be significantly more than SR predictions.

\subsection{Doppler Shift}
The Doppler shift is one of the few dynamic properties of a celestial
object that we can measure directly.  The measured redshift is easily
translated to a speed, yielding a view of an expanding universe.  As
shown in Appendix~\ref{sec.z}, the redshift $1+\z$ depends on the real
and apparent speeds as follows:
\begin{eqnarray}
\nonumber
1 \,+\, \z \,&=&\, \frac{1}{1 \,+\, \bp\cos\theta}\\
\nonumber
 \,&=&\, 1 \,-\,\beta\cos\theta
\end{eqnarray}
where $\beta$ is the real speed of the object, and $\bp$ is its
apparent speed. For a receding object ($\theta=\pi$) moving at
subluminal speeds ($\beta<1$), we can rewrite this equation as:
\begin{equation}
\nonumber
1 \,+\, \z \,=\, \sqrt{\frac{1+\beta}{1 \,-\,
      \bp}}
\end{equation}
If we were to mistakenly assume that the speed we observe is the real
speed, then this becomes the familiar relativistic Doppler shift
formula:
\begin{equation}
\nonumber
1 \,+\, \z \,=\, \sqrt{\frac{1+\beta}{1 \,-\,
      \beta}}
\end{equation}
While it is interesting that we get the relativistic Doppler shift
formula, it should be noted that setting $\bp = \beta$ breaks down
the derivation of these equations.  However, this similarity in the
forms of the final equations is indicative of the common basis in
their origin.

\section{LTT Effects for Approaching Objects}
\subsection{Asymmetric Effects}
One important feature of LTT effects is that they are asymmetric in
their dependence on speed; $\beta$ and $\bp$ appear in odd power, so
that the equations are odd.  More generally, there is a term involving
the angle $\theta$ between the object's velocity and the observer's
line of sight.  In SR, on the other hand, $\beta$ almost always
appears as $\beta^2$ and the equations are even.  SR treats the effect
of motion as a linear coordinate transformation, ignoring the angle.
Thus, in SR, the effect is the same whether the object is approaching
or receding from the observer.  As remarked before, this fundamental
difference can be traced back to the assumed homogeneity of space and
time in SR.  The asymmetry in LTT effects, on the other hand, provides
convincing explanations to certain paradoxes: the twin paradox, the
observed superluminal motion, the causality violation due to
superluminal motion etc.  At the same time, the asymmetry makes it
difficult to reconcile LTT effects and SR completely.

\subsection{Time Contraction and Length Expansion}
The asymmetric consequences of LTT effects include an apparent time
contraction. When an object is approaching the observer, the time at
the object seems to flow at an accelerated rate for the observer.
This effect is easy to understand because, as the object is
approaching the observer, the successive photons are emitted at
shorter and shorter distances and they take less and less time to
reach the observer, creating an illusion of an accelerated time flow,
or time contraction.

By the same argument, the moving object appears elongated along the
direction of motion as it is flying towards the observer.
Appendices~\ref{sec.tp} and \ref{sec.lp} show the mathematical details
of how light travel time effects result in an apparent time
contraction and length expansion.  If $\Delta\ta$ is the apparent time
duration as felt by the observer and $\Delta t$ is the real time,
then:
\begin{eqnarray}
\nonumber
\frac{\Delta\ta}{\Delta t} &\,=\,& \frac{1}{1+\bp}
\end{eqnarray}
Similarly, an object of real length $d$ appears to have an elongated
length $\da$ as given by:
\begin{eqnarray}
\nonumber
\frac{\da}{d} &\,=\,& {1+\bp}
\end{eqnarray}

\section{Explanations Based on Light Travel Time Effect}

\subsection{Twin Paradox}
The famous twin paradox in SR exploits the symmetry in its coordinate
transformation.  In this paradox, one twin goes away to a galaxy far
away, accelerating to speeds close to $c$.  The other one stays back
on earth.  When the traveling twin comes back (again accelerating to
almost $c$ on the way), he will be much younger than the twin that
stays back, due to time dilation.  But, in the traveling twin's frame
of reference, it is the other twin (along with the earth) that is
traveling at speeds close to $c$.  Thus, time dilation should apply to
the one that stays back.  This paradox is resolved by arguing that the
traveling twin feels the tremendous acceleration and deceleration, and
his frame of reference is not an inertial frame.

In the LTT picture, the time dilation equation is asymmetric.  Whatever
time dilation one twin seems to feel on the way out is compensated by
an exactly same amount of contraction on his way back.  Thus, to each
of the twins, the other twin seems to be enjoying the benefits of time
dilation and aging slower.  But, this time dilation happens only
during the outward journey, when the twins are going away from each
other. On his way back, the traveling twin will see the other twin
aging much faster.  At the same time, to the twin that stays back, the
traveling twin will appear to be aging much faster.  When they meet
again, there will not be any age difference.

\subsection{Superluminality and Causality}
Although superluminality is generally believed to lead to time travel
and the consequent causality violations, SR does not explicitly state
this.  As quoted earlier, Einstein merely remarked that ``our
deliberations become meaningless'' at superluminal speeds. In any
case, we saw that for a receding object, the apparent speed could
never be superluminal.  And SR considers only receding frames of
reference.  In our derivations of LTT effects on length contraction
and time dilation, we did not impose the condition that $\beta<1$.

Using our equation for time dilation, for an approaching object,
$\theta = 0$.
\begin{eqnarray}
\nonumber
  \frac{\Delta\ta}{\Delta t}  &\,=\,& {1-\beta}
\end{eqnarray}
Thus, if the object is flying to the observer, up to the speed of
light ($0 < \beta < 1$), the time intervals appear shorter and
shorter.  When the speed of approach exceeds $c$, the {\em apparent}
time flows backwards.  This is because a photon emitted at a
particular point along the trajectory reaches the observer {\em
  before} a photon emitted earlier and farther away.  The order in
which photons emitted by the object reach the observer is reversed.
This reversal of time flow will give rise to an apparent violation of
causality.  This violation of causality is only an LTT effect (akin to
a video clip playing backwards), not a fundamental property of space
and time as currently believed.  Note, however, that astrophysical
causality violations may not be obvious. For instance, imagine a
cataclysmic explosion of a star and a subsequent fireball. This
scenario played backwards would be a imploding fireball and an
appearance of a star.  We may think of it as the accretion of matter
by an invisible massive object or the birth of a star, instead of an
event showing causality violation.

\subsection{Apparent Superluminal Motion}
\image{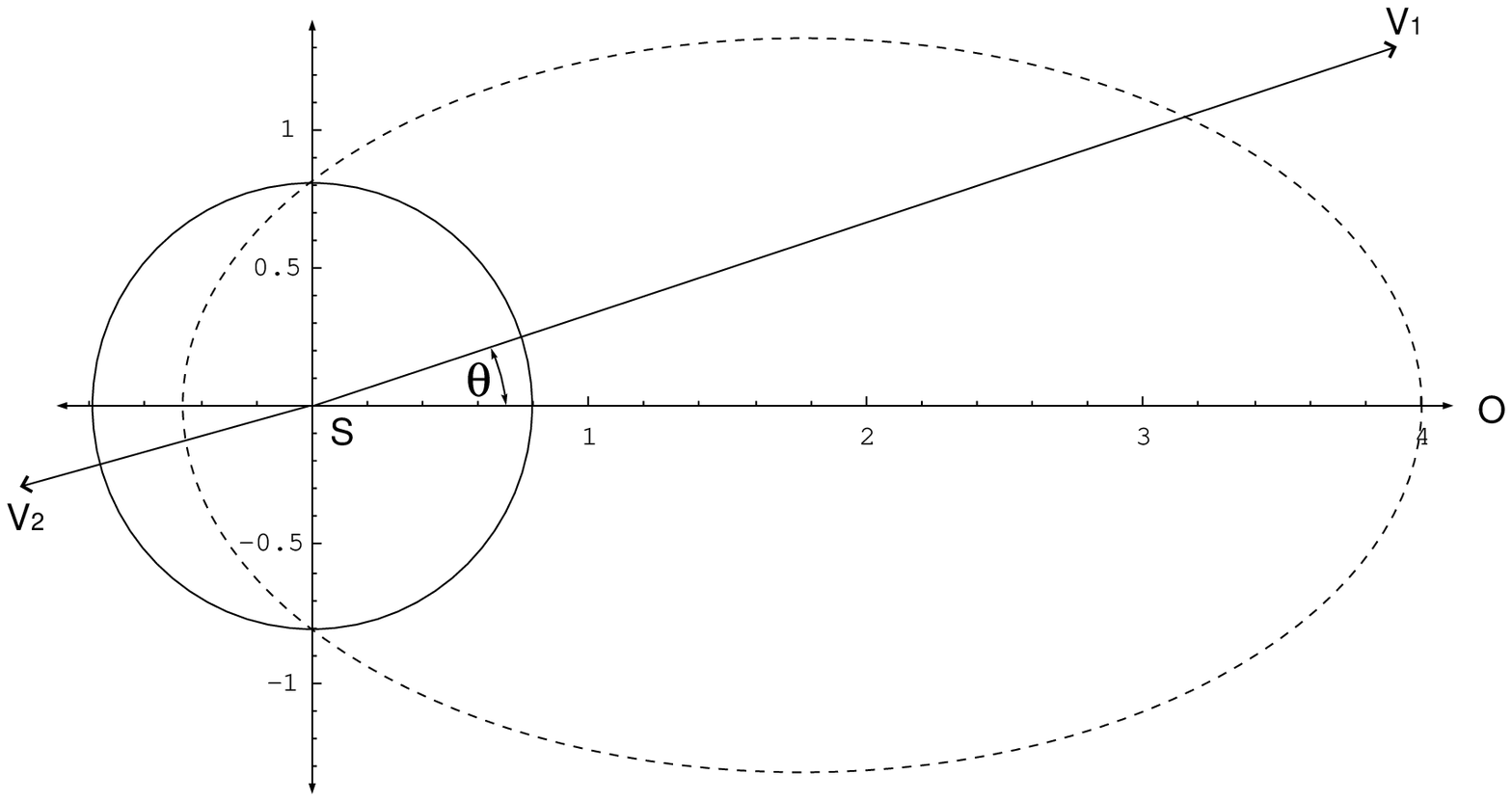}{8}{rees1} {Illustration of the traditional
  explanation for the apparent superluminal motion.  An object
  expanding at a speed $\beta = 0.8$, starting from a single point S.
  The solid circle represents the boundary one second later.  The
  observer is far away on the right hand side, O ($x\to\infty$).  The
  dashed ellipse is the apparent boundary of the object, as seen by
  the observer.}

We can measure the transverse velocity of a celestial object almost
directly using angular measurements, which are translated to a speed
using its known (or estimated) distance from us.  In the past few
decades, scientists have observed \citep{M87,superluminal2} objects
moving at apparent transverse velocities significantly higher than the
speed of light.  Some such superluminal objects were detected within
our own galaxy
\citep{superluminal1,superluminal3,superluminal4,GRS1915}.
\citet{rees} offered an explanation why such apparent superluminal
motion is not in disagreement with SR based on LTT effects, even
before the phenomenon was discovered.

The distortion in the perception of speed, when the object is
approaching the observer, is used to explain the apparent superluminal
motion.  \Figref{rees1} illustrates the explanation of apparent
superluminal motion as described in the seminal paper by \citet{rees}.
In this figure, the object at S is expanding radially at a constant
speed of $0.8c$, a highly relativistic speed.  The part of the object
expanding along the direction $V_1$, close to the line of sight of the
observer, will appear to be traveling much faster.  This will result
in an apparent transverse velocity that can be superluminal.

Imagine an object in motion at a speed $\beta$.  To an observer, it
appears to move with a speed of $\bp$. The apparent speed $\bp$ of the
object depends on the real speed $\beta$ and the angle between its
direction of motion and the observer's line of sight, $\theta$.  As
shown in Appendix~\ref{sec.bp},
\begin{equation}
\bp \,=\, \frac{\beta}{1\,-\,\beta \cos\theta} \label{eqn.1}
\end{equation}

\Figref{rees1} is a representation of \eqnref{eqn.1} as
$\cos\theta$ is varied over its range.  It is the locus of $\bp$
for a constant $\beta = 0.8$, plotted against the angle $\,\theta$.
The apparent speed is in complete agreement with what was predicted in
1966 (Figure~1 in the original article, \citet{rees}).

For a narrow range of $\,\theta$, the transverse component of the
apparent velocity ($\,\bp\sin\theta\,$) can appear superluminal.
From \eqnref{eqn.1}, it is easy to find this range:
\begin{equation}
\frac{1 - \sqrt{2 \beta^2-1}}{2\beta} < \cos\theta < \frac{1 +
  \sqrt{2 \beta^2-1}}{2\beta} \label{eqn.2}
\end{equation}
Thus, for appropriate values of $\,\beta (>\frac{1}{\sqrt{2}})\,$ and
$\,\theta\,$ (as given in \eqnref{eqn.2}), the transverse
velocity of an object can seem superluminal, even when the real
speed is in conformity with the special theory of relativity.

While \eqnsref{eqn.1}{eqn.2} explain the apparent transverse
superluminal motion the difficulty arises in the recessional side.
Along directions such as $V_2$ in \figref{rees1}, the apparent
velocity is always smaller than the real velocity.  It can be shown
that the apparent velocity of the slower jet can never be more than
the reciprocal of the faster jet, if the real speeds are to be
subluminal.  This calculation is shown in Appendix~{sec:jet}. Thus,
superluminality can never be observed in both the jets of a radio
source, which indeed has not been reported so far.  Near exact
symmetry in extragalactic radio sources, including subluminal jets, is
also qualitatively inconsistent with this explanation.

\subsection{Symmetric Radio Sources}


If we accept that special relativity applies to our cognitive map of
reality or the perceived space and time, and that the absolute
reality, of which space and time are our perception, is free of the
constraints of SR, we can find elegant descriptions of symmetric radio
sources and jets.  Different classes of such objects associated with
Active Galactic Nuclei (AGN) were found in the last fifty years.  The
radio galaxy \astrobj{Cygnus A} \citep{cyga}, one of the brightest
radio objects, is an example of such a radio source.  Many of its
features are common to most extragalactic radio sources: the symmetric
double lobes, an indication of a core, an appearance of jets feeding
the lobes and the hotspots.  \citet{hotspot1} and \citet{hotspot2}
have reported more detailed kinematical features, such as the proper
motion of the hotspots in the lobes.  Here, we show that our
perception of an object crossing our field of vision at a constant
superluminal speed is remarkably similar to a pair of symmetric
hotspots departing from a fixed point with a decelerating rate of
angular separation.

\image{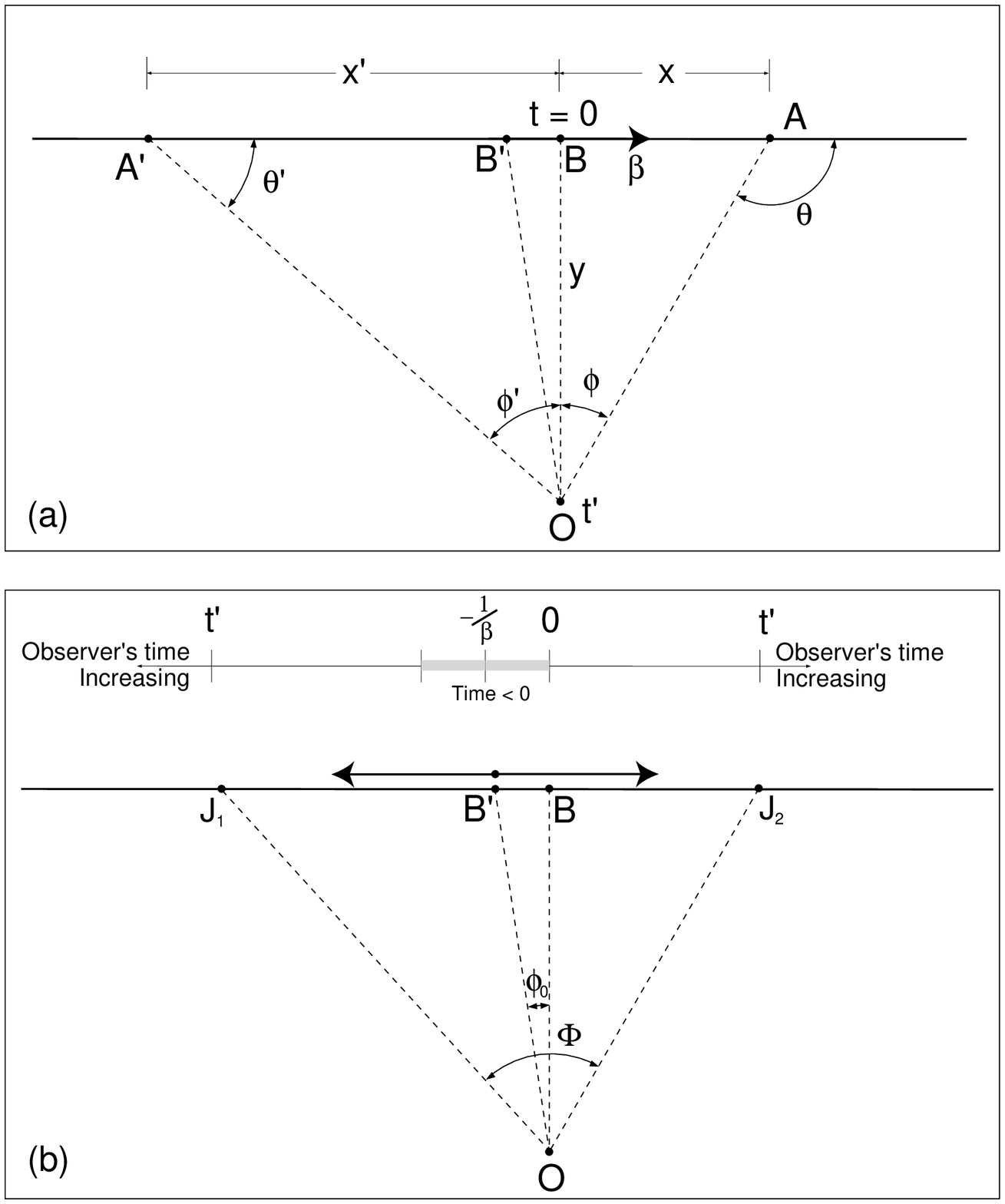}{8}{d} {The top panel (a) shows an object flying along
  $A_-BA$ at a constant superluminal speed.  The observer is at $O$.
  The object crosses $B$ (the point of closest approach to $O$) at
  time $t=0$.  The bottom panel (b) shows how the object is perceived
  by the observer at $O$.  It first appears at $B'$, then splits into
  two.  The two apparent objects seem to go away from each other
  (along $J_1$ and $J_2$) as shown.}

Consider an object moving at a superluminal speed as shown in
\figref{d}(a).  The point of closest approach is $B$.  At that
point, the object is at a distance of $y$ from the observer at $O$.
Since the speed is superluminal, the light emitted by the object at
some point $B'$ (before the point of closest approach $B$) reaches the
observer {\em before\/} the light emitted at $A_-$.  This reversal
creates an illusion of the object moving in the direction from $B'$ to
$A_-$, while in reality it is moving in the opposite direction.

We use the variable $\ta$ to denote the observer's time.  Note
that, by definition, the origin in the observer's time axis is set
when the object appears at $B$.  $\phi$ is the observed angle with
respect to the point of closest approach $B$. $\phi$ is defined as
$\theta - \pi/2$ where $\theta$ is the angle between the object's
velocity and the observer's line of sight.  $\phi$ is negative for
negative time $t$.

As shown in Appendix~\ref{sec.super}, a relation between $\ta$ and
$\phi$ can be readily derived.
\begin{equation}
 \ta = y\left( \frac{\tan\phi}\beta + \frac{1}{\cos\phi} - 1\right)
 \label{eqn.3}
\end{equation}
Here, we have chosen units such that $c = 1$, so that $y$ is also the
time light takes to traverse $BO$.  The origin of the observer's time
is set when the observer sees the object at $B$.  \ie $\ta = 0$ when
the light from the point of closest approach $B$ reaches the observer.

The actual plot of $\phi$ as a function of the observer's time is
given in \figref{PhiVsTp0} for different speeds $\beta$.  Note that
for subluminal speeds, there is only one angular position for any
given $\ta$.  For subluminal objects, the observed angular position
changes almost linearly with the observed time, while for superluminal
objects, the change is parabolic.  The time axis scales with $y$.

\Eqnref{eqn.3} can be approximated using a Taylor series
expansion as:
\begin{equation}
\ta \approx y\left(\frac\phi\beta +
  \frac{\phi^2}{2}\right)\label{eqn.4}
\end{equation}
From the quadratic \eqnref{eqn.4}, one can easily see that the
minimum value of $\ta$ is $\ta_\text{min} = -y/2\beta^2$ and it occurs
at $\phi_{0}=-1/\beta$.  Thus, to the observer, the object first
appears (as though out of nowhere) at the position $\phi_0$ at time
$\ta_\text{min}$.  Then it appears to stretch and split, rapidly at
first, and slowing down later.

The angular separation between the objects flying away from each other
is:
\begin{equation}
\nonumber
 \Phi = \frac{2}{\beta}\sqrt{1+\frac{2\beta^2}{y}\ta} =
 \frac{2}{\beta}\left(1+\beta\phi\right)
\end{equation}
And the rate at which the separation occurs is:
\begin{equation}
\nonumber
 \frac{d\Phi}{d\ta} = \sqrt{\frac{2}{y t_\text{age}}} =
 \frac{2\beta}{y\left(1+\beta\phi\right)} \label{eqn.5}
\end{equation}
where $ t_\text{age} = \ta - \ta_\text{min}$, the apparent age of the
symmetric object. (The mathematical details can be found in
Appendix~\ref{sec.super}.)

\image{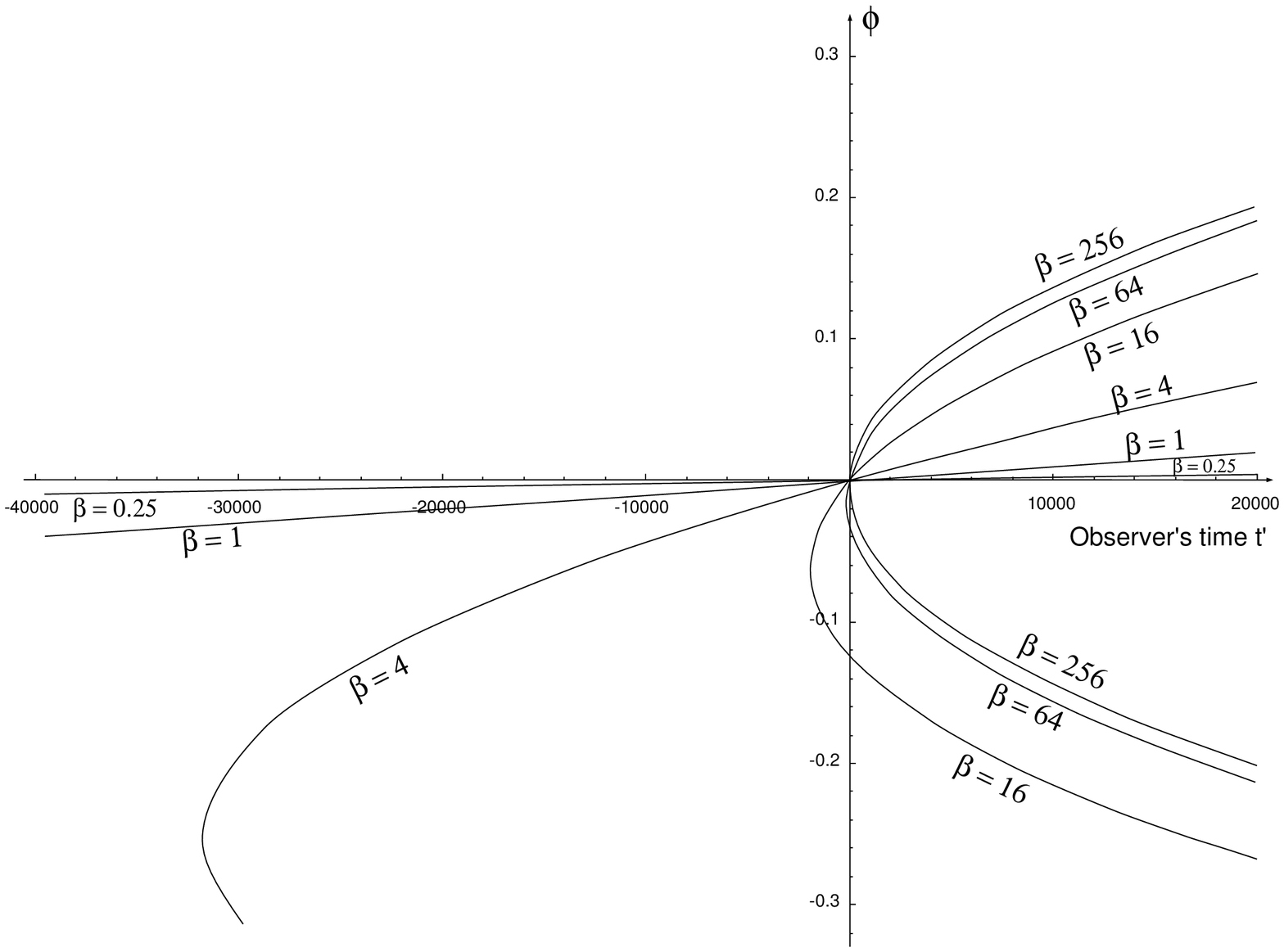}{8}{PhiVsTp0} {The apparent angular positions of an
  object traveling at different speeds at a distance $y$ of one
  million light years from us.  The angular positions ($\phi$ in
  radians) are plotted against the observer's time $\ta$ in years.}

This discussion shows that a single object moving across our field of
vision at superluminal speed creates an illusion of an object
appearing at a at a certain point in time, stretching and splitting
into two and then moving away from each other.  This time evolution of
the two objects is given in \eqnref{eqn.3}, and illustrated in
the bottom panel of \figref{d}(b).  Note that the apparent time
$\ta$ (as perceived by the observer) is reversed with respect to the
real time $t$ in the region $A_-$ to $B'$.  An event that happens near
$B'$ appears to happen before an event near $A_-$.  Thus, the observer
may see an apparent violation of causality, but it is just a part of
the light travel time effect.

If there are multiple objects, moving as a group, at roughly constant
superluminal speed along the same direction, they will appear as
a series of objects materializing at the same angular position and moving
away from each other sequentially, one after another.  The apparent
knot in one of the jets always has a corresponding knot in the other
jet.

\subsection{Redshifts of the Hotspots}
In the previous section, we showed how a superluminal object appears
as two objects receding from a core.  Now we consider the time
evolution of the redshift of the two apparent objects (or hotspots).
Since the relativistic Doppler shift equation is not appropriate for
our considerations, we need to work out the relationship between the
redshift ($\z$) and the speed ($\beta$) from first principles.  This
calculation is done in Appendix~\ref{sec.z}:
\begin{eqnarray}
\nonumber
1 \,+\, \z \,&=&\, \left|1 \,-\, \beta\cos\theta\right| \\
\nonumber
\,&=&\, \left|1  \,+\, \beta\sin\phi\right|\\
\,&=&\, \left|1 \,+\, \frac{\beta^2t}{\sqrt{\beta^2t^2 + y^2}}\right|
\label{eqn.z0}
\end{eqnarray}

We can explain the radio frequency spectra of the hotspots as
extremely redshifted black body radiation, because $\beta$ can be very
large in our model of extragalactic radio sources.  Note that the
limiting value of $|1+\z|$ is roughly $\beta$, which gives an
indication of the speeds required to push the black body radiation to
RF spectra. Since the speeds ($\beta$) involved are typically very
large, and we can approximate the redshift as:
\begin{equation}
\nonumber
1 \,+\, \z \,\approx\,  \left|\beta\phi\right| \,\approx\,
\frac{\left|\beta\Phi\right|}{2}
\end{equation}
Assuming the object to be a black body similar to the sun, we can
predict the peak wavelength (defined as the wavelength at which the
luminosity is a maximum) of the hotspots as:
\begin{equation}
\nonumber
\lambda_\text{max} \,\approx\, (1+\z) 480nm \,\approx\,
\frac{\left|\beta\Phi\right|}{2} 480nm \label{eqn.z1}
\end{equation}
where $\Phi$ is the angular separation between the two hotspots.

This equation shows that the peak RF wavelength increases linearly
with the angular separation.  If multiple hotspots can be located in a
twin jet system, their peak wavelengths will depend only on their
angular separation, in a linear fashion.  Such a measurement of the
emission frequency as $\phi$ increases along the jet is clearly seen
in the photometry of the jet in \astrobj{3C 273} \citep{RF2UV}.
Furthermore, if the measurement is done at a single wavelength,
intensity variation can be expected as the hotspot moves along the
jet.  In other words, measurements at higher wavelengths will find the
peak intensities farther away from the core region, which is again
consistent with observations.

\subsection{Gamma Ray Bursts}

The evolution of redshift of the thermal spectrum of a superluminal
object also holds the explanation for gamma ray bursts (GRBs).
$\gamma$ ray bursts are short and intense flashes of $\gamma$ rays in
the sky, lasting from a few milliseconds to several minutes
\citep{GRB1}.  The short flashes (the prompt emissions) are followed
by an after-glow of progressively softer energies. Thus, the initial
$\gamma$ rays are promptly replaced by X-rays, light and even radio
frequency waves. This softening of the spectrum has been known for
quite some time \citep{GRB5}, and was first described using a
hypernova (fireball) model.  In this model, a relativistically
expanding fireball produces the $\gamma$ emission, and the spectrum
softens as the fireball cools down \citep{piran2}.  The model
calculates the energy released in the $\gamma$ region as
$10^{53}$--$10^{54}$ ergs in a few seconds.  This energy output is
similar to about 1000 times the total energy released by the sun over
its entire lifetime.

More recently, an inverse decay of the peak energy with varying time
constant has been used to empirically fit the observed time evolution
of the peak energy \citep{GRB3,GRB4} using a collapsar model.  According
to this model, GRBs are produced when the energy of highly
relativistic flows in stellar collapses are dissipated, with the
resulting radiation jets angled properly with respect to our line of
sight.  The collapsar model estimates a lower energy output, because
the energy release is not isotropic, but concentrated along the jets.
However, the rate of the collapsar events has to be corrected for the
fraction of the solid angle within which the radiation jets can appear
as GRBs.  GRBs are observed roughly at the rate of once a day.  Thus,
the expected rate of the cataclysmic events powering the GRBs is of
the order of $10^4$--$10^6$ per day. Because of this inverse
relationship between the rate and the estimated energy output, the
total energy released per observed GRB remains the same.

Symmetric radio sources (galactic or extragalactic) and GRBs may
appear to be completely distinct phenomena.  However, their cores show
a similar time evolution in the peak energy, but with very different
time constants.  Other similarities have begun to attract attention in
the recent years \citep{GRB7}. Treating GRB as a manifestation of the
light travel time results in a model that unifies these two phenomena
and makes detailed predictions of their kinematics.

\image{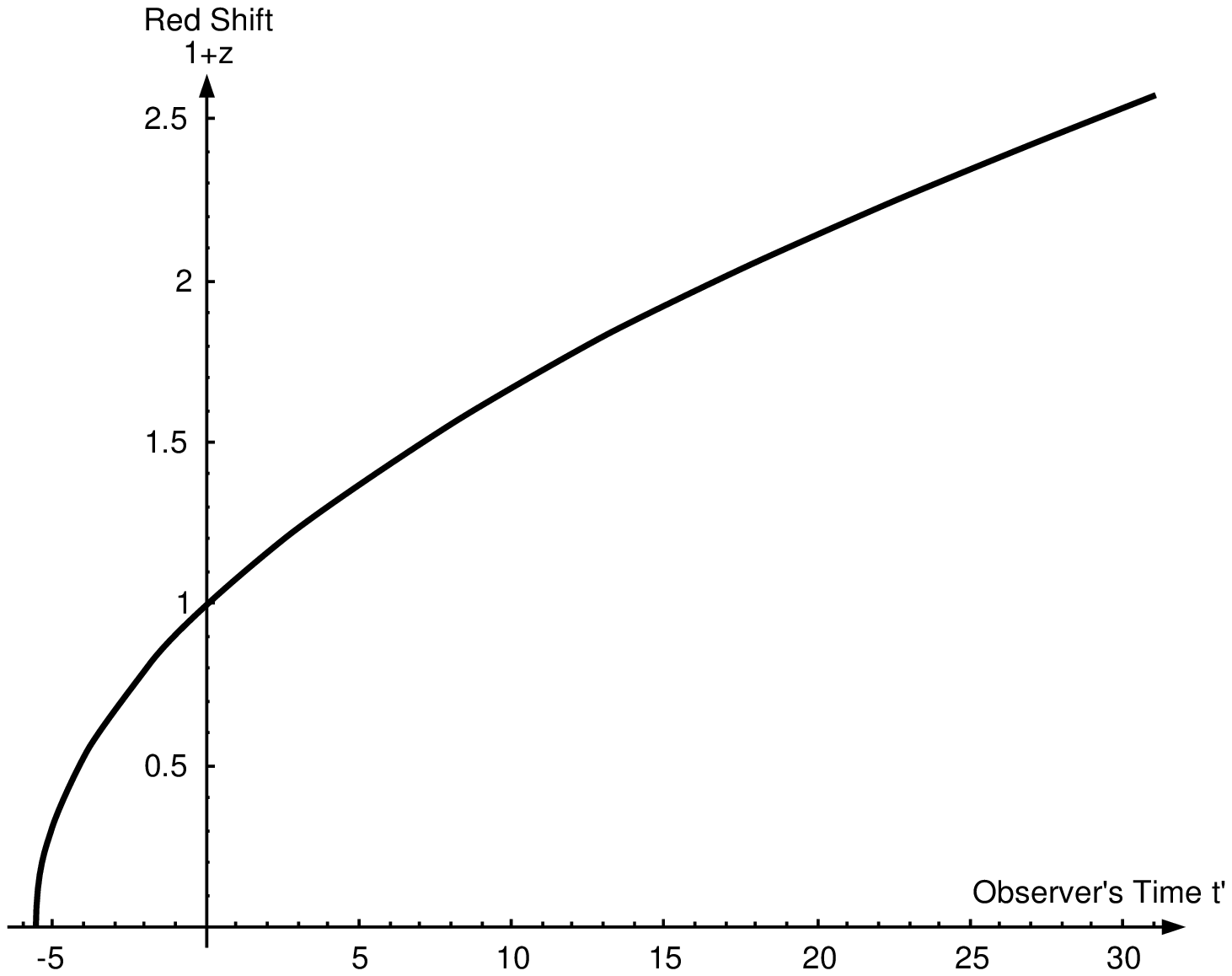}{8}{zVsTp}{Time evolution of the redshift from a
  superluminal object.  It shows the redshifts
  expected from an object moving at $\beta = 300$ at a distance of ten
  million light years from us.  The X axis is the observer's time in
  years.  (Since the X axis scales with time, it is also the redshift
  from an object at 116 light days --ten million light seconds--
  with the X axis representing $\ta$ in seconds.)}

The spectra of GRBs rapidly evolve from $\gamma$ region to an optical
or even RF after-glow.  This evolution is similar to the spectral
evolution of the hotspots of a radio source as they move from the core
to the lobes.  The evolution of GRB can be made quantitative, because
we know the dependence of the observer's time $\ta$ and the redshift
$1+\z$ on the real time $t$ (\eqnsref{eqn.3}{eqn.z0}).  From these
two, we can deduce the observed time evolution of the redshift (see
Appendix~\ref{sec.dz}).  We have plotted it parametrically in
\figref{zVsTp} that shows the variation of redshift as a function of
the observer's time ($\ta$).  The figure shows that the observed
spectra of a superluminal object is expected to start at the
observer's time $\ta_\text{min}$ with heavy (infinite) blue shift.
The spectrum of the object rapidly softens and soon evolves to zero
redshift and on to higher values.  The rate of softening depending on
its speed and distance from us.  The speed and the distance are the
only two parameters that are different between GRBs and symmetric
radio sources in our model.

Note that the X axis in \figref{zVsTp} scales with time.  We have
plotted an object with $\beta = 300$ and $y =$ ten million light
years, with X axis is $\ta$ in years.  It is also the variation of
$1+\z$ for an object at $y =$ ten million light seconds (or 116 light
days) with X axis in seconds.  The former corresponds to symmetric
jets and the latter to a GRB.  Thus, for a GRB, the spectral evolution
takes place at a much faster pace.  Different combinations of
$\beta$ and $y$ can be fitted to describe different GRB spectral
evolutions.

To the observer, there is no object before $\ta_\text{min}$.  In
other words, there is a definite point in the observer's time when the
GRB is ``born'', with no indication of its impending birth before that
time.  This birth does not correspond to any cataclysmic event (as
would be required in the collapsar/hypernova or the ``fireball''
model) at the distant object.  It is just an artifact of our
perception.

In order to compare the time evolution of the GRB spectra to the ones
reported in the literature, we need to get an analytical expression
for the redshift ($\z$) as a function of the observer's time ($\ta$).
This can be done by eliminating $t$ from the equations for $\ta$ and
$1+\z$ (\eqnsref{eqn.3}{eqn.z0}), with some algebraic manipulations as
shown in Appendix~\ref{sec.dz}.  The algebra can be made more
manageable by defining $\tau = y/\beta$, a characteristic time scale
for the GRB (or the radio source).  This is the time the object would
take to reach us, if it were coming directly toward us.  We also
define the age of the GRB (or radio source) as $t_\text{age} = \ta -
\ta_\text{min}$.  This is just the observer's time ($\ta$) shifted by
the time at which the object first appears to him ($\ta_\text{min}$).
With these notations (and for small values $t$), it is possible to
write the time dependence of $\z$ as:
\begin{equation}
  \label{eqn.z}
1 + \z = \left| {1 + \frac{{\beta ^2 \left( { - \tau  \pm \sqrt
            {2 \beta  t_{\text{age}}} } \right)}}{{\beta t_{\text{age}}
        + \tau /2 \mp \sqrt {2 \beta t_{\text{age}}}  + \beta ^2 \tau
      }}} \right|
\end{equation}
for small values of $t \ll \tau$.

Since the peak energy of the spectrum is inversely proportional to the
redshift, it can be written as:
\begin{equation}
  \label{eqn.Epk0}
  E_\text{pk}(t_{\text{age}}) = \frac{ E_\text{pk}(\ta_\text{min})}{1 + C_1
  \,\sqrt{\frac{t_{\text{age}}}{\tau}} + C_2\,\frac{t_{\text{age}}}{\tau}}
\end{equation}
where $C_1$ and $C_2$ are coefficients to be estimated by the Taylor
series expansion of \eqnref{eqn.z} or by fitting.

\citet{GRB6} have studied the evolution of the peak energy
($E_\text{pk}(t)$), and modeled it empirically as:
\begin{equation}
  \label{eqn.Epk}
  E_\text{pk}(t) = \frac{ E_\text{pk,0}}{(1+t/\tau)^\delta}
\end{equation}
where $t$ is the time elapsed after the onset ($= t_\text{age}$ in our
notation), $\tau$ is a time constant and $\delta$ is the hardness
intensity correlation (HIC).  \citet{GRB6} reported seven fitted
values of $\delta$.  We calculate their average as $\delta =
1.038\pm0.014$, with the individual values ranging from $0.4$ to
$1.1$.  Although it may not rule out or validate either model within
the statistics, the $\delta$ reported may fit better to
\eqnref{eqn.Epk0}.  Furthermore, it is not an easy fit, because there
are too many unknowns.  However, the similarity between the shapes of
\eqnsref{eqn.Epk0}{eqn.Epk} is remarkable, and points to the agreement
between our model and the existing data.

\subsection{Expansion of the Universe}
\label{sec.universe}
Our perception of superluminal motion also leads to the appearance of
an expanding universe. The expansion of the universe is inferred by
the redshift measurements of recessional speeds. The apparent
recessional speed is the longitudinal component of $\bp$ is
$\bp_\parallel = \bp \cos\theta$.  From \eqnref{eqn.1}, we can see
that
\begin{eqnarray}
\nonumber
\bp_\parallel \,=\,\bp \cos\theta\,&=&\, \frac{\beta \cos\theta}{1\,-\,\beta \cos\theta} \\
\nonumber
\lim_{\beta\to\pm\infty} \bp_\parallel \,&=&\, -1
\end{eqnarray}
The apparent recessional speed tends to $c$ (or,
$\,\bp_\parallel\,\to\,-1$), when the real speed is highly
superluminal.  This limiting value of $\bp_\parallel$ is independent
of the actual direction of motion of the object $\theta$.  Thus,
whether a superluminal object is receding or approaching (or, in fact,
moving in any other direction), its appearance from our perspective
will be that of an object receding from us roughly at the speed of
light.

The recessional speeds are measured using redshifts that, by
\eqnref{eqn.z2}, tend to large values as $\,\bp_\parallel\,\to\,-1$.
\begin{eqnarray}
\nonumber
1 \,+\, \z \,&=&\, \frac{1}{1 \,+\, \bp\cos\theta}\\
 \,&=&\, \frac{1}{1 \,+\, \bp_\parallel}
\label{eqn.z2}
\end{eqnarray}
Thus, the appearance of all (possibly superluminal) objects receding from
us at strictly subluminal speeds is an artifact of our perception,
rather than the true nature of the universe.

\subsection{Cosmic Microwave Background Radiation}

The red shift of celestial objects $1+\z$ also has an interesting
limiting value at large angles, and for superluminal speeds.
\begin{eqnarray}
\nonumber
1+\z \,&=&\,  |1 \,+\, \beta\sin\phi| \\
\nonumber
\lim_{\phi\to\pm\pi/2} 1 + \z  \,&=&\, |1+\beta| \,\approx\, \beta
\end{eqnarray}
Thus, if we picture our universe as a large number of superluminal or
hyperluminal objects moving around in random directions, there will be
a significant amount of low energy isotropic electromagnetic
radiation.  A low energy isotropic spectrum is remarkably similar to
the cosmic microwave background radiation (CMBR).  Thus, CMBR can be
explained if we think of our visual reality as being limited by the
light travel time effects.  Note than it is not just our perception
that gets fooled by the LTT effects, our measurement instruments also
work at the speed of light and are subject to the same constraints.

\section{Conclusions}

In this article, we started with an insight from cognitive
neuroscience about the nature of reality.  Reality is a convenient
representation that our brain creates out of our sensory inputs.  This
representation, though convenient, is an incredibly distant
experiential mapping of the actual physical causes that make up the
inputs to our senses.  Furthermore, limitations in the chain of
sensing and perception map to measurable and predictable
manifestations to the reality we perceive.  One such fundamental
constraint to our perceived reality is the speed of light, and the
corresponding manifestations are generally termed the light travel
time (LTT) effects.  Because space and time are a part of a reality
created out of light inputs to our eyes, some of their properties are
manifestations of LTT effects, especially on our perception of motion.
The absolute, physical reality generating the light inputs does not
obey the properties we ascribe to our perceived space and time.

Noting that SR only considers frames of reference receding from each
other, we showed that LTT effects are qualitatively identical.  This
similarity is not surprising because the coordinate transformation in
SR is derived based partly on light travel time effects, and partly on
the assumption that light travels at the same speed with respect to
all inertial frames. In treating it as a manifestation of LTT, we did
not address the primary motivation of SR, which is a covariant
formulation of Maxwell's equations, as evidenced by the opening
statements of Einstein's original paper \citep{einstein}. It may be
possible to disentangle the covariance of electrodynamics from the
coordinate transformation, although it is not attempted in this
article.

Unlike SR, LTT effects are asymmetric.  This asymmetry provides a
resolution to the twin paradox and an interpretation of the assumed
causality violations associated with superluminality.  Furthermore,
the perception of superluminality is modulated by LTT effects, and
explains $\gamma$ ray bursts and symmetric jets.  As we showed in the
article, perception of superluminal motion also holds an explanation
for cosmological phenomena like the expansion of the universe and
cosmic microwave background radiation.  The light travel time effects
should be considered as a fundamental constraint in our perception,
and consequently in physics, rather than as a convenient explanation
for isolated phenomena.

Given that our perception is filtered through LTT effects, we have to
deconvolute them from our perceived reality in order to understand the
nature of the absolute, physical reality.  This deconvolution,
however, results in multiple solutions.  Thus, the absolute, physical
reality is beyond our grasp, and any {\em assumed} properties of the
absolute reality can only be validated through how well the resultant
{\em perceived} reality agrees with our observations.  In this
article, we assumed that the {\em absolute} reality obeys our
intuitively obvious classical mechanics and asked the question how
such a reality would be perceived when filtered through light travel
time effects. We demonstrated that this particular treatment could
explain certain astrophysical and cosmological phenomena that we
observe.

The coordinate transformation in SR is a redefinition of space and
time (or, more generally, reality) in order to accommodate the
distortions in our perception of motion due to light travel time
effects.  The absolute reality behind our perception is not subject to
restrictions of SR.  One may be tempted to argue that SR applies to
the ``real'' space and time, not our perception.  This line of
argument begs the question, what is real?  Reality is nothing but a
cognitive model created in our brain starting from our sensory inputs,
visual inputs being the most significant.  Space itself is a part of
this cognitive model.  Perceptual constraints map directly to the
nature of space as we perceive it.  We have no access to a reality
beyond our perception.  The choice of accepting the perception of
reality as a true image of reality and redefining space and time as
described in special relativity indeed amounts to a philosophical
choice.  The alternative presented in the article is prompted by the
view in modern neuroscience that reality is a cognitive model in the
brain based on our sensory inputs.  Adopting this alternative reduces
us to guessing the nature of the absolute reality and comparing its
predicted projection to our real perception.  It may simplify and
elucidate some theories in physics and explain some puzzling phenomena
in our universe.  However, this option is just another philosophical
stance against the unknowable absolute reality.

\appendix

\section{Mathematical Details}

\image{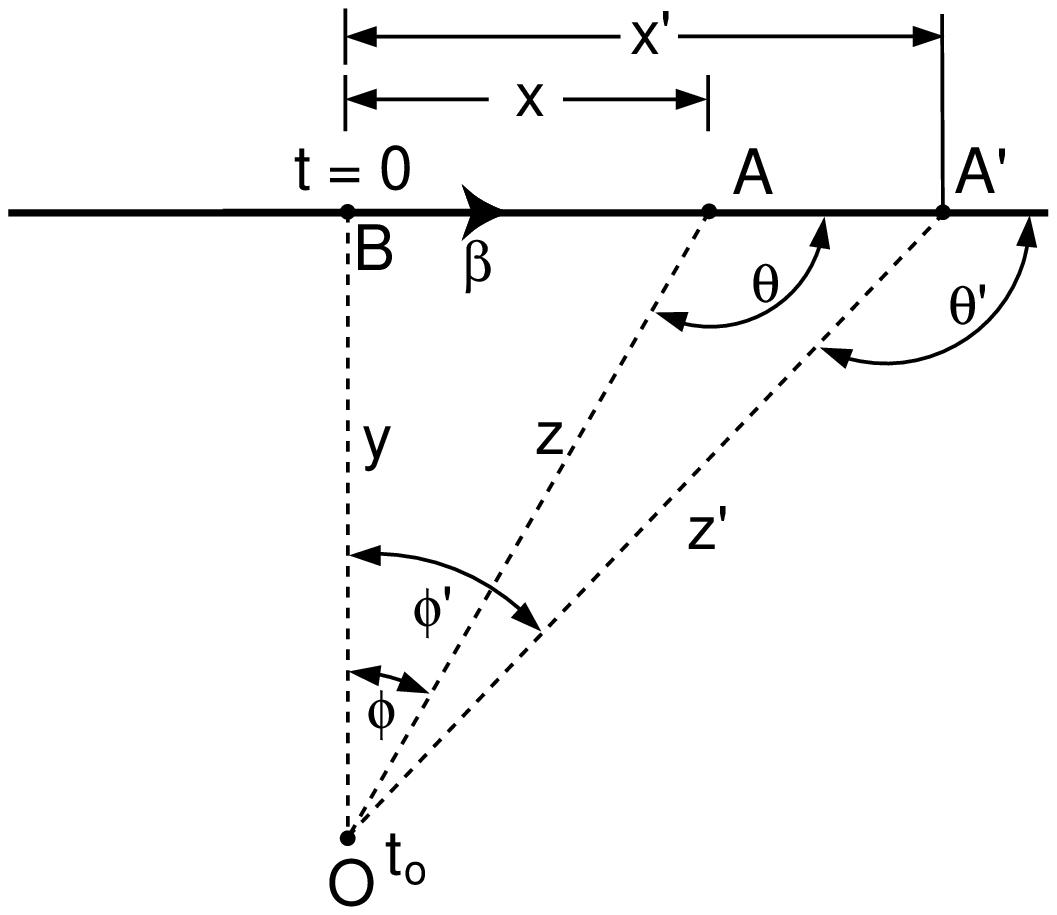}{8}{c} {The object is flying along $BAA'$, the observer
  is at $O$.  The object crosses $B$ (the point of closest approach) at
  time $t=0$.  It reaches $A$ at time $t$.  A photon emitted at $A$
  reaches $O$ at time $\ta$, and a photon emitted at $A'$ reaches $O$
  at time $\ta'$.}

\subsection{Perception of Speed}
\label{sec.bp}
In this section, we derive how the perception of speed is distorted
due to the light travel time (LTT) effects.  We will show that the
apparent speed is limited to the speed of light when the object is
receding from us.

In \figref{c}, there is an observer at $O$. An object is flying by
at a high speed $v = \beta c$ along the horizontal line $BAA'$.  With
no loss of generality, we can assume that $t = 0$ when the object is
at $B$, the point of closest approach. It passes $A$ at time $t$. The
photon emitted at time $t = 0$ reaches the observer at time $t = \ta$,
and the photon emitted at $A'$ (at time $t = t'$) reaches him at time
$t = \ta'$.  The angle between the object's velocity at $A$ and the
observer's line of sight is $\theta$.  We have the Pythagoras
equations:
\begin{eqnarray}
\nonumber
z^2 \,&=&\, x^2 \,+\, y^2 \\
\nonumber
z'^2 \,&=&\, x'^2 \,+\, y^2\\
\Rightarrow\qquad \frac{x+x'}{z+z'} &\,=\,& \frac{z-z'}{x-x'} \label{eqn.6}
\end{eqnarray}
If we assume that $x$ and $z$ (distances at time $t_0$) are not very
different from $x'$ and $z'$ respectively (distances at time $\ta$), we
can write,
\begin{equation}
\label{eqn.cos}
-\cos\theta = \sin\phi = \frac{x}{z} \approx \frac{x'+x}{z'+z} =
\frac{z'-z}{x'-x}
\end{equation}
We define the real speed of the object as:
\begin{equation}
\label{eqn.b}
v \,=\, \beta\,c \,=\, \frac{x'\,-\,x}{t'-t}
\end{equation}
But the speed it {\em appears\/} to have will depend on when the
observer senses the object at $A$ and $A'$.  The apparent speed of the
object is:
\begin{equation}
\label{eqn.bp}
v' \,=\, \bp\,c \,=\, \frac{x' \,-\, x}{\ta' \,-\,\ta}
\end{equation}
We also have
\begin{eqnarray}
\nonumber
\ta \,&=&\, t+\frac{z}{c}\\
\nonumber
 \ta'\,&=&\, t' + \frac{z'}{c}\\
\Rightarrow \ta'-\ta \,&=&\, t' - t + \frac{z'-z}{c}
\end{eqnarray}
Thus,
\begin{eqnarray}
\nonumber
\frac{\beta}{\bp} \,&=&\, \frac{\ta'-\ta}{t'-t} \\
\nonumber
\,&=&\, 1 + \frac{z'-z}{c(t'-t)} \\
\nonumber
\,&=&\, 1 - \frac{x-x'}{c(t'-t)}\cos\theta \\
\,&=&\, 1 - \beta\cos\theta
\end{eqnarray}
which gives,
\begin{eqnarray}
\nonumber
\bp &\,=\,&\frac{\beta}{1\,-\,\beta \cos\theta}\\
\beta &\,=\,&\frac{\bp}{1\,+\,\bp \cos\theta}
\end{eqnarray}
and,
\begin{eqnarray}
\nonumber
 \frac{\bp}{\beta}  &\,=\,& \frac{1}{1-\beta\,\cos\theta} \\
\nonumber
&\,=\,& 1+\bp\,\cos\theta\\
&\,=\,& \sqrt{\frac{1+\bp\,\cos\theta}{1-\beta\,\cos\theta}}
\label{eqn.7}\end{eqnarray}

LTT effects modulate the way we perceive time at objects in motion.
Here we show that a receding object appears to have a dilated time
flow. From \figref{c}, we can see that $\theta = \pi$ for an object
receding from the observer.  Thus, the apparent speed of a receding
object is:
\begin{eqnarray}
\nonumber
\bp &\,=\,& \frac{\beta}{1+\beta} \\
\lim_{\beta\to\pm\infty} \bp &\,=\,& 1
\end{eqnarray}
Thus, an object can never appear to be receding faster than the speed
of light.

\subsection{Time Dilation}
\label{sec.tp}
Referring to \figref{c}, we can see that the real time elapsed as
the object moves from $A$ to $A'$ is:
\begin{equation}
  \Delta t = t' - t
\end{equation}
This time period appears to the observer as:
\begin{equation}
  \Delta\ta = \ta' - \ta
\end{equation}
Using the definitions of the real and apparent speeds as in
\eqnsref{eqn.b}{eqn.bp}, we can write:
\begin{eqnarray}
\nonumber
  \frac{\Delta\ta}{\Delta t} &\,=\,& \frac{\beta}{\bp} \\
\nonumber
 &\,=\,& {1-\beta\,\cos\theta} \\
&\,=\,& \frac{1}{1+\bp\,\cos\theta}
\end{eqnarray}
where we used the known relationship between $\beta$ and $\bp$ from
\eqnref{eqn.7}.

For an object receding from the observer, $\theta=\pi$ and the
equation becomes:
\begin{eqnarray}
  \frac{\Delta\ta}{\Delta t} &\,=\,& \frac{1}{1-\bp}
\end{eqnarray}

For an object approaching the observer, $\theta=0$ and the
equation becomes:
\begin{eqnarray}
  \frac{\Delta\ta}{\Delta t} &\,=\,& \frac{1}{1+\bp}
\end{eqnarray}
This shows a time contraction, instead of a time dilation.

\subsection{Length Contraction}
\label{sec.lp}
\image{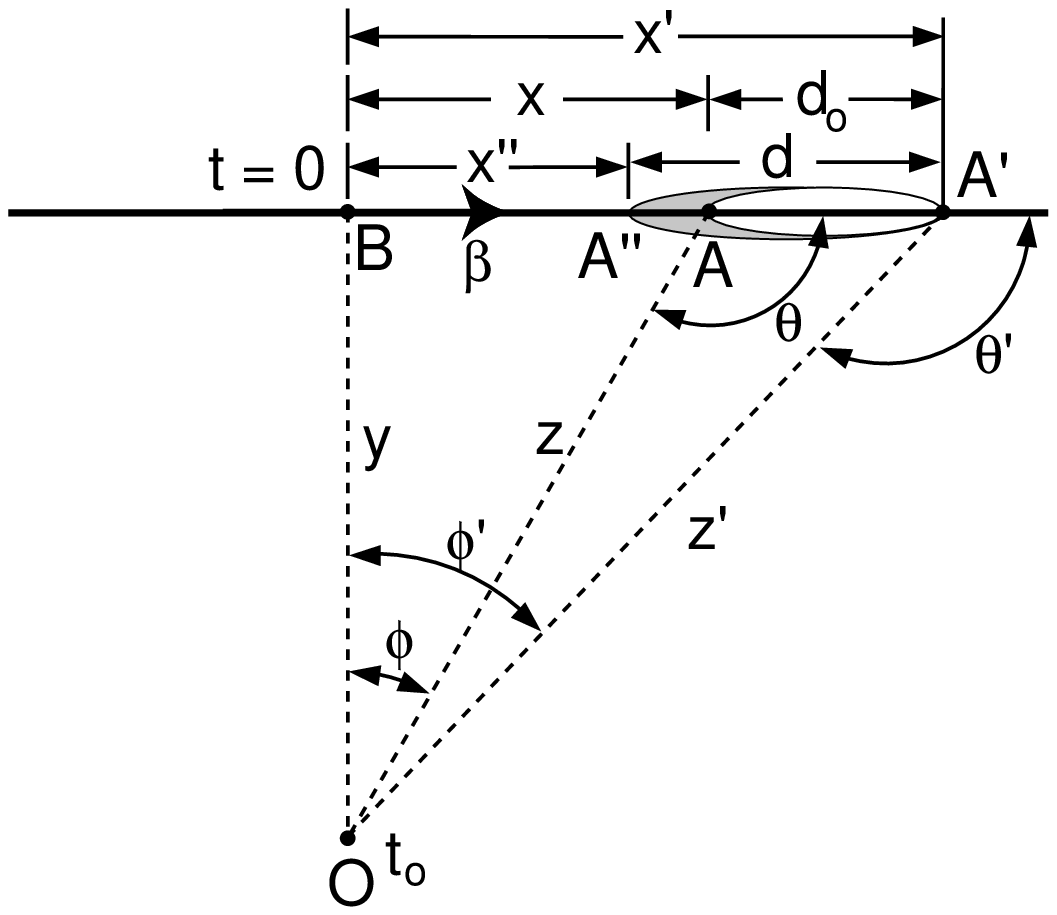}{8}{c2} {The object has a real length of $d$, and is shown
  as the shaded ellipse.  To the observer at $O$, it appears to have a
  length of $\da$ due to LTT effects.}

The perceived length of an object in motion is affected due to LTT
effects.  In particular, a receding object appears shorter. In
\figref{c2}, we have the object of real length $d$.  The perceived
length of the object is the distance between the leading edge and the
trailing edge from which the photons reach the observer at the same
instant.  In \figref{c2}, it is denoted by $\da$.  The photon emitted
from the trailing edge of the object when it is at $x$ reaches the
observer at $O$ at time $\ta$.  At the same time, a photon from the
leading edge at $x'$ reaches $O$.  But, when the leading edge is at
$x'$, the trailing edge is only at $x'' = x'-d$, due to the motion.

Since the object's speed is $v$ and the time starts when the object
passes $B$, we can write:
\begin{eqnarray}
\nonumber
  \ta \,&=&\, \frac{x}{v} + \frac{z}{c}\\
\nonumber
\,&=&\, \frac{x''}{v} + \frac{z'}{c}\\
\,&=&\, \frac{x'-d}{v} + \frac{z'}{c}
\end{eqnarray}
Using the equation for the apparent length of the object $\da = x' -
x$, we can rewrite this as:
\begin{eqnarray}
\nonumber
  \frac{z'-z}{c} \,&=&\, \frac{x-x'+d}{v} \\
\,&=&\, \frac{d-\da}{v}
\label{eqn.len}
\end{eqnarray}
The approximation for $\cos\theta$ in \eqnref{eqn.cos} is
still valid, with the additional information that the apparent length
of the object $\da = x' - x$.
\begin{equation}
  -\cos\theta = \frac{z'-z}{\da}
\end{equation}
Thus, \eqnref{eqn.len} becomes:
\begin{equation}
    \label{eqn.len1}
-\da\,\cos\theta = \frac{d-\da}{\beta}
\end{equation}
Or,
\begin{eqnarray}
\nonumber
\frac{\da}{d} &\,=\,& \frac{1}{1-\beta\,\cos\theta} \\
\nonumber
&\,=\,& {1+\bp\,\cos\theta} \\
&\,=\,& \frac{\bp}{\beta}
\label{eqn.len2}
\end{eqnarray}

For an object receding from the observer, $\theta=\pi$ and the
equation becomes:
\begin{eqnarray}
\frac{\da}{d} &\,=\,& {1-\bp}
\end{eqnarray}

For an object approaching the observer, $\theta=0$ and the
equation becomes:
\begin{eqnarray}
\frac{\da}{d} &\,=\,& {1+\bp}
\end{eqnarray}
which shows that the apparent length of the object is greater than its
real length.

\subsection{Doppler Shift}
\label{sec.z}
Redshift ($\z$) defined as:
\begin{equation}
1 \,+\, \z \,=\, \frac{\lp}{\lambda}
\end{equation}
where $\lp$ is the measured wavelength and $\lambda$ is the known
wavelength.  In \figref{c}, the number of wave cycles created in
time $t'-t$ between $A$ and $A'$ is the same as the number of wave
cycles sensed at $O$ between $\ta'$ and $\ta$.  Substituting the
values, we get:
\begin{equation}
\frac{(t'-t)\, c}{\lambda} \,=\, {\frac{(\ta'\,-\,\ta)\,c}
{\lp}}
\end{equation}
Using the definitions of the real and apparent speeds
from \eqnsref{eqn.b}{eqn.bp}, it is easy to get:
\begin{equation}
\frac{\lp}{\lambda} \,=\, \frac{\beta}{\bp}
\end{equation}
Using the relationship between the real speed $\beta$ and the
apparent speed $\bp$ from \eqnref{eqn.7},
we get:
\begin{eqnarray}
\nonumber
1 \,+\, \z \,&=&\, \frac{1}{1 \,+\, \bp\cos\theta}\\
 \,&=&\, 1 \,-\,\beta\cos\theta
\end{eqnarray}
As expected, $\z$ depends on the longitudinal component of the
velocity of the object.  Since we allow superluminal speeds in this
calculation, we need to generalize this equation for $\z$ noting that
the ratio of wavelengths is positive.  Taking this into account, we
get:
\begin{eqnarray}
\nonumber
1 \,+\, \z \,&=&\, \left|\frac{1}{1 \,+\,
      \bp\cos\theta}\right| \\
\,&=&\, \left|1 \,-\, \beta\cos\theta\right| \label{eqn.8}
\end{eqnarray}
For a receding object $\theta=\pi$.  If we consider only subluminal
speeds, we can rewrite this as:
\begin{equation}
1 \,+\, \z \,=\, \sqrt{\frac{1+\beta}{1 \,-\,
      \bp}}
\end{equation}
If we were to mistakenly assume that the speed we observe is the real
speed, then this becomes the relativistic Doppler formula:
\begin{equation}
1 \,+\, \z \,=\, \sqrt{\frac{1+\beta}{1 \,-\,
      \beta}}
\end{equation}
\subsection{Kinematics of Superluminal Objects}
\label{sec.super}
\image{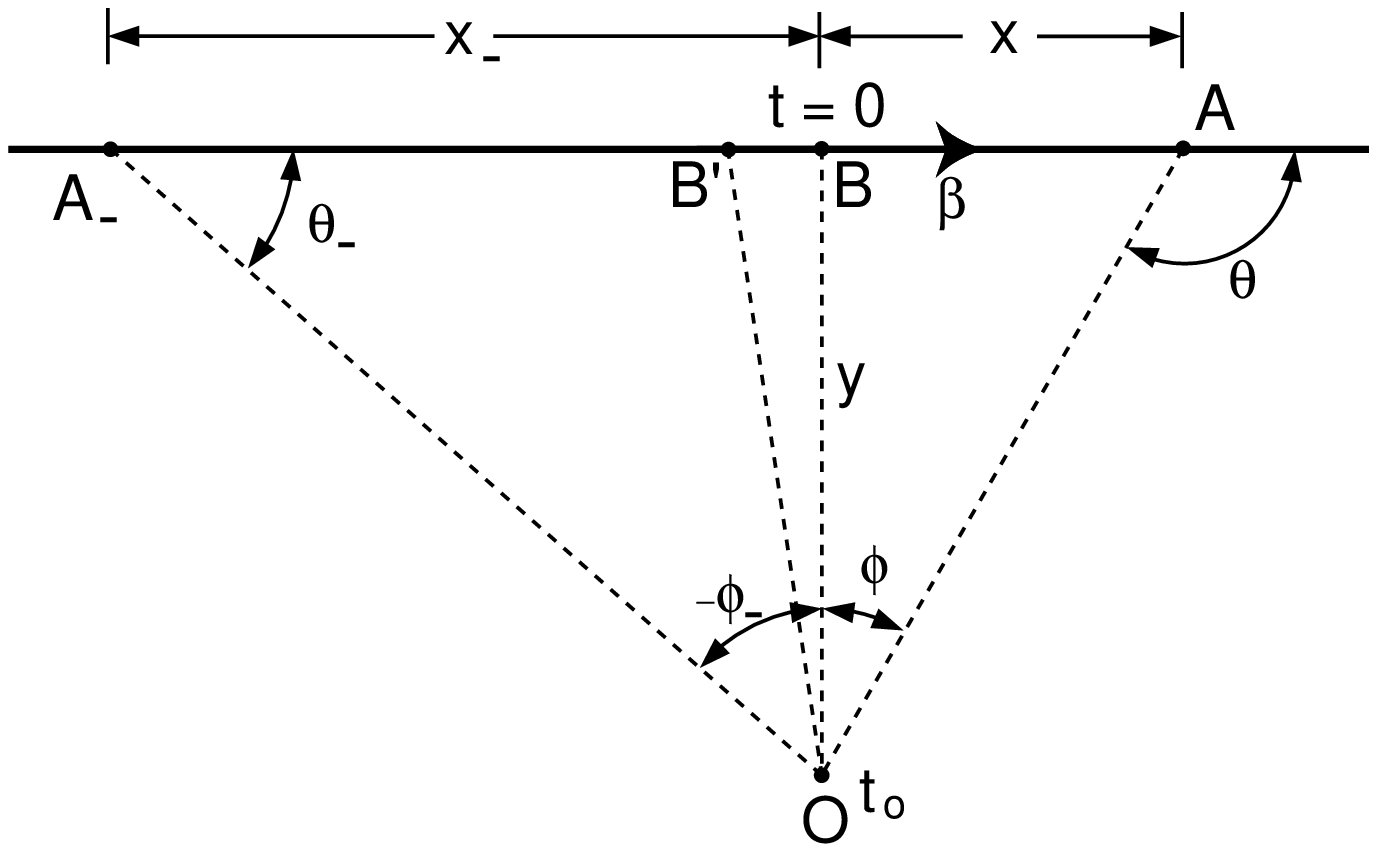}{8}{dd} {An object flying along $A_-BA$ at a constant
  superluminal speed.  The observer is at $O$.  The object crosses $B$
  (the point of closest approach to $O$) at time $t=0$.}

The derivation of the kinematics is based on \figref{dd}.  Here, an
object is moving at a superluminal speed along $A_-BA$. At the point of
closest approach, $B$, the object is a distance of $y$ from the
observer at $O$. Since the speed is superluminal, the light emitted by
the object at some point $B'$ (before the point of closest approach
$B$) reaches the observer {\em before\/} the light emitted at $A_-$.
This gives an illusion of the object moving in the direction from $B'$
to $A_-$, while in reality it is moving from $A_-$ to $B'$.

$\phi$ is the observed angle with respect to the point of closest
approach $B$.  $\phi$ is defined as $\theta - \pi/2$ where $\theta$ is
the angle between the object's velocity and the observer's line of
sight.  $\phi$ is negative for negative time $t$.  We choose units
such that $c = 1$, in order to make algebra simpler.  $\ta$ denotes
the observer's time.  Note that, by definition, the origin in the
observer's time, $\ta$ is set to the instant when the object appears
at $B$.

The real position of the object at any time $t$ is:
\begin{equation}
 x = y\tan\phi = \beta t
\end{equation}
Or,
\begin{equation}
 t = \frac{y\tan\phi}{\beta}
\end{equation}
A photon emitted by the object at $A$ (at time $t$) will reach $O$
after traversing the hypotenuse.  A photon emitted at $B$ will reach
the observer at $t = y$, since we have chosen $c = 1$.  If we define
the observer's time $\ta$ such that the time of arrival is $t = \ta +
y$, then we have:
\begin{equation}
 \ta = t + \frac{y}{\cos\phi} - y
\end{equation}
which gives the relation between $\ta$ and $\phi$.
\begin{equation}
 \ta = y\left( \frac{\tan\phi}\beta + \frac{1}{\cos\phi} - 1\right)
\end{equation}
Expanding the equation for $\ta$ to second order, we get:
\begin{equation}
 \ta = y\left(\frac\phi\beta + \frac{\phi^2}{2}\right)\label{eqn.9}
\end{equation}
The minimum value of $\ta$ occurs at $\phi_{0}=-1/\beta$ and it is
$\ta_\text{min} = -y/2\beta^2$.  To the observer, the object first
appears at the position $\phi=-1/\beta$.  Then it appears to stretch
and split, rapidly at first, and slowing down later.

The quadratic \eqnref{eqn.9} can be recast as:
\begin{equation}
  \label{eqn.q1}
  1+\frac{2\beta^2}{y}\ta = \left(1+\beta\phi\right)^2
\end{equation}
which will be more useful later in the derivation.

The angular separation between the objects flying away from each other
is the difference between the roots of the quadratic
\eqnref{eqn.9}:
\begin{eqnarray}
\nonumber
 \Phi \,&=&\, \phi_1-\phi_2 \\
\nonumber
\,&=&\, \frac{2}{\beta}\sqrt{1+\frac{2\beta^2}{y}\ta} \\
\,&=&\, \frac{2}{\beta}\left(1+\beta\phi\right)
\end{eqnarray}
making use of \eqnref{eqn.q1}.  Thus, we have the angular separation
either in terms of the observer's time ($\Phi(\ta)$) or the angular
position of the object ($\Phi(\phi)$) as illustrated in
Figure~\ref{phiphi}.

\image{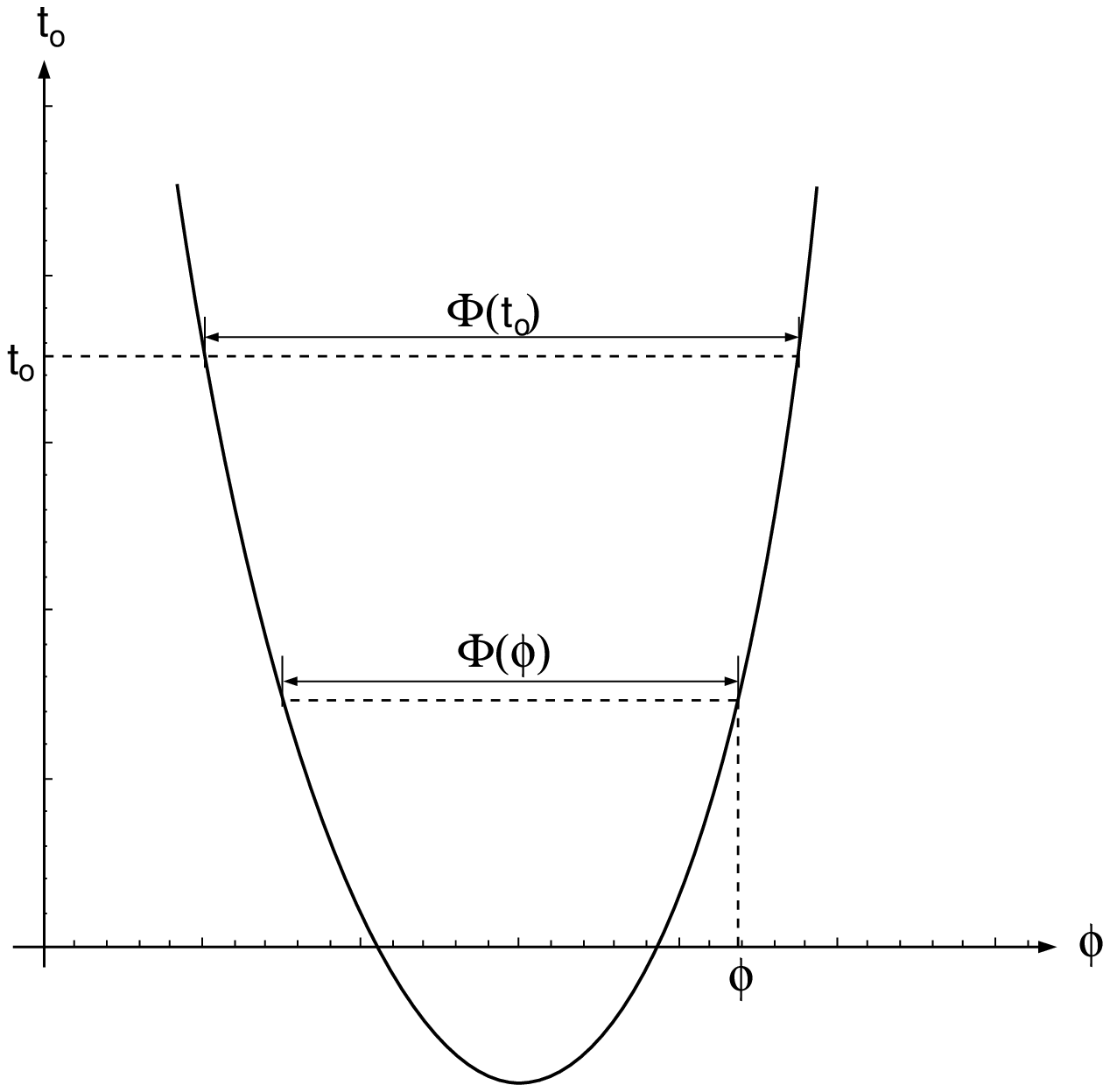}{8}{phiphi} {Illustration of how the angular separation
  is expressed either in terms of the observer's time ($\Phi(\ta)$) or
  the angular position of the object ($\Phi(\phi)$)}

The rate at which the angular separation occurs is:
\begin{eqnarray}
\nonumber
 \frac{d\Phi}{d\ta} \,&=&\, \frac{2\beta}{y\sqrt{1+\frac{2\beta^2}{y}\ta}}
 \\
\,&=&\, \frac{2\beta}{y\left(1+\beta\phi\right)}
\end{eqnarray}
Again, making use of \eqnref{eqn.q1}.  Defining the apparent age of
the radio source $ t_\text{age} = \ta - \ta_\text{min}$ and knowing
$\ta_\text{min} = -y/2\beta^2$, we can write:
\begin{eqnarray}
\nonumber
 \frac{d\Phi}{d\ta} \,&=&\, \frac{2\beta}{y\sqrt{1+\frac{2\beta^2}{y}\ta}}\\
\nonumber
\,&=&\, \frac{2\beta}{y\sqrt{1-\frac{\ta}{\ta_\text{min}}}}\\
\nonumber
\,&=&\, \sqrt{\frac{4\beta^2}{y^2}\,\times\,\frac{-\ta_\text{min}}{\ta-\ta_\text{min}}}\\
\,&=&\,\sqrt{\frac{2}{y\, t_\text{age}}}
\end{eqnarray}

\subsection{Time Evolution of the Redshift}
\label{sec.dz}
As shown before in \eqnref{eqn.8}, the redshift $\z$ depends
on the real speed $\beta$ as:
\begin{equation}
1 \,+\, \z \,=\, \left|1 \,-\, \beta\cos\theta\right| \,=\, \left|1
  \,+\, \beta\sin\phi\right|\label{eqn.10}
\end{equation}
For any given time ($\ta$) for the observer, there are two solutions
for $\phi$ and $\z$.  $\phi_1$ and $\phi_2$ lie on either side of
$\phi_0 = 1/\beta$.  For $\sin\phi > -1/\beta$, we get
\begin{equation}
 1+\z_2 = 1+\beta\sin\phi_1
\end{equation} and for $\sin\phi < -1/\beta$,
\begin{equation}
 1+\z_1 = -1 - \beta\sin\phi_2
\end{equation}
Thus, we get the difference in the redshift between the two hotspots
at $\phi_1$ and $\phi_2$ as:
\begin{equation}
 \Delta\z \approx 2 + \beta(\phi_1+\phi_2)
\end{equation}
We also have the mean of the solutions of the quadratic ($\phi_1$ and
$\phi_2$) equal to the position of the minimum ($\phi_0$):
\begin{equation}
\frac{\phi_1 + \phi_2}{2} = -\frac{1}{\beta}
\end{equation}
Thus $\phi_1+\phi_2 = -2/\beta$ and hence $\Delta\z = 0$.  The two
hotspots will have identical redshifts, if terms of $\phi^3$ and above
are ignored.

As shown before (see \eqnref{eqn.10}), the redshift $\z$
depends on the real speed $\beta$ as:
\begin{equation}
1 \,+\, \z \,=\,\left|1 \,+\, \beta\sin\phi\right| \,=\, \left|1 \,+\,
  \frac{\beta^2t}{\sqrt{\beta^2t^2 + y^2}}\right|
\end{equation}
Since we know $\z$ and $\ta$ functions of $t$, we can plot their
inter-dependence parametrically.  This is shown in \figref{zVsTp} of
the article.

It is also possible to eliminate $t$ and derive the dependence of
$1+\z$ on the apparent age of the object under consideration,
$t_\text{age} = \ta - t_\text{min}$.  In order to do this, we first
define a time constant $\tau = y/\beta$.  This is the time the object
would take to reach us, if it were flying directly toward us.  First,
let's get an expression for $t/\tau$:
\begin{eqnarray}
\nonumber
 \ta \,&=&\, t + \sqrt {\beta ^2 t^2  + y^2 }  - y \\
\nonumber
  \,&=&\, t + \beta \tau \sqrt {1 + \frac{{t^2 }}{{\tau ^2 }}}  - \beta \tau  \\
\nonumber
  \,&\approx&\, t + \frac{{\beta t^2 }}{{2\tau }} \\
  \Rightarrow \frac{t}{\tau } \,&=&\, \frac{{ - 1 \pm \sqrt {1 +
  \frac{{2\beta t_\text{age} }}{\tau }} }}{\beta }
\label{eqn.ttau}
\end{eqnarray}
Note that this is valid only for $t \ll \tau$.  Now we collect the
terms in $t/\tau$ in the equation for $1+\z$:
\begin{eqnarray}
\nonumber
 \ta \,&=&\, t + \sqrt {\beta ^2 t^2  + y^2 }  - y \\
\nonumber
 \Rightarrow \sqrt {\beta ^2 t^2  + y^2 }  \,&=&\, \ta - t + y \\
\nonumber
 1 + z \,&=&\, \left| {1 + \frac{{\beta ^2 t}}{{\sqrt {\beta ^2 t^2  + y^2 } }}} \right| \\
\nonumber
  \,&=&\, \left| {1 + \frac{{\beta ^2 t}}{{\ta - t + y}}} \right| \\
  \,&=&\, \left| {1 + \frac{{\beta ^2 \frac{t}{\tau
 }}}{{\frac{{t_{\text{age}} }}{\tau } - \frac{1}{{2\beta }} - \frac{t}{\tau } + \beta }}}
 \right|
\label{eqn.zz}
\end{eqnarray}
As expected, the time variables always appear as ratios like $t/\tau$,
giving confidence that our choice of the characteristic time scale is
probably right.  Finally, we can substitute $t/\tau$ from
\eqnref{eqn.ttau} in \eqnref{eqn.zz} to obtain:
\begin{equation}
1 + \z = \left| {1 + \frac{{\beta ^2 \left( { - \tau  \pm \sqrt
            {2 \beta  t_{\text{age}}} } \right)}}{{\beta t_{\text{age}}
        + \tau /2 \mp \sqrt {2 \beta t_{\text{age}}}  + \beta ^2 \tau
      }}} \right|
\end{equation}

\subsection{Estimating Real Speed from Apparent Speed}
\label{sec:jets}

\image{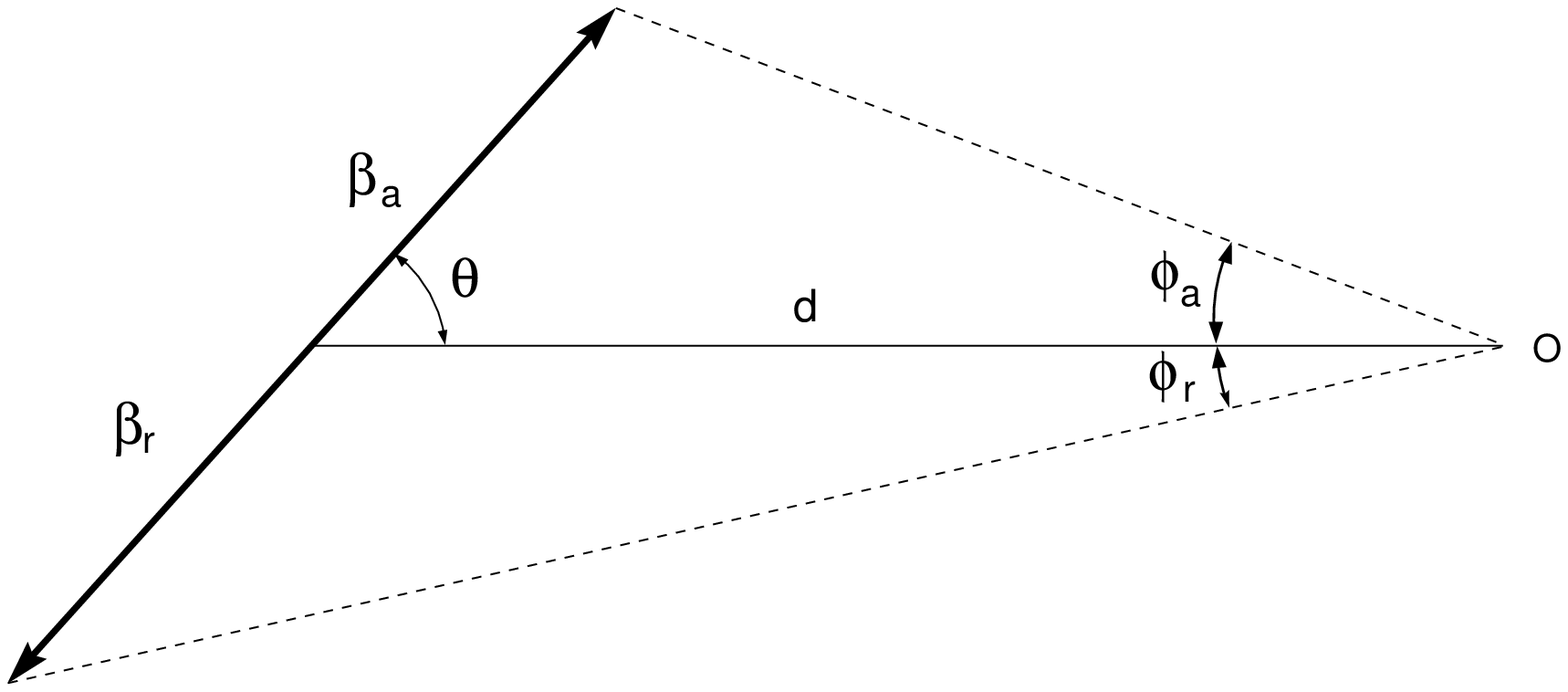}{8}{jets} {Illustration of the real jet speeds ($\beta_a$
  and $\beta_r$), core distance ($d$) and the angles.}

In the traditional explanation of superluminality, superluminal
objects such as \astrobj{GRS 1915+105} are assumed to be two jets
emanating from a core.  The axis of the jets makes an angle $\theta$
with respect to our line of sight.  The only direct kinematic
measurements we have are the angular velocities of features (or knots)
in the jets.  We have two angular rates, $\mu_a$ and $\mu_r$, for the
approaching and receding jets.  The distance of the core from us ($d$)
is not known.  Also unknown are the real speeds of the jets $\beta_a$
and $\beta_r$, which are usually assumed to be the same ($\beta$).
The apparent transverse speeds ($\bp_\perp^a$ and $\bp_\perp^r$) are
different for the two jets. Thus, we have the following definitions:
\begin{eqnarray}
\mu_a  \,&=&\, \frac{d\phi_a}{d\ta}\\
\mu_r  \,&=&\, \frac{d\phi_r}{d\ta} \\
\bp_\perp^a  \,&=&\, \mu_a\,d \\
\bp_\perp^r  \,&=&\, \mu_r\,d
\end{eqnarray}
where $\ta$ is our time.  Assuming the real jet speeds are the same
($\beta_a = \beta_r = \beta$) and using the relationship between
$\beta$ and $\bp$ from \eqnref{eqn.7}, we have the following
equations:
\begin{eqnarray}
\label{eqn:bpp1}
   \mu_a\,d \,&=&\, \frac{\beta\sin\theta}{1-\beta\cos\theta}\\
\label{eqn:bpp2}
   \mu_r\,d \,&=&\, \frac{\beta\sin\theta}{1+\beta\cos\theta}
\end{eqnarray}
There are three unknowns ($\beta, \theta$ and $d$) and only two
equations. Thus, it is always possible to impose the relativistic
condition ($\beta < 1$) and compute corresponding limits on $\theta$
and $d$. The only way to estimate the real speed or the angle is to
have an independent (and, hopefully, model-independent) measurement of
$d$.

In order to find the limiting values of $\bp_\perp^a$ and
$\bp_\perp^r$, we set $\beta\to1$ in \eqnsref{eqn:bpp1}{eqn:bpp2}.
\begin{eqnarray}
   \bp_\perp^a  \,&=&\, \frac{\sin\theta}{1-\cos\theta}\\
   \bp_\perp^r  \,&=&\, \frac{\sin\theta}{1+\cos\theta}
\end{eqnarray}
Or,
\begin{eqnarray}
  \bp_\perp^a  \,&=&\, \frac{\sin\theta}{1-\cos\theta}\\
  \,&=&\, \frac{\sqrt{1-\cos^2\theta}}{1-\cos\theta} \\
  \,&=&\, \frac{\sqrt{(1-\cos\theta)(1+\cos\theta)}}{1-\cos\theta}\\
  \,&=&\, \sqrt{\frac{1+\cos\theta}{1-\cos\theta}} \\
  \,&=&\, \sqrt{\frac{(1+\cos\theta)(1+\cos\theta)}{(1-\cos\theta)(1+\cos\theta)}} \\
  \,&=&\, \frac{1+\cos\theta}{\sqrt{1-\cos^2\theta}} \\
  \,&=&\, \frac{1+\cos\theta}{\sin\theta} \\
  \,&=&\, \frac{1}{\bp_\perp^r}
\end{eqnarray}
Thus, if we assume that the real speeds are limited to $\beta < 1$,
the apparent transverse speed of the receding jet ($\bp_\perp^r$) is
limited to the reciprocal of the apparent transverse speed of the
approaching jet ($\bp_\perp^a$).  As long as the measured angular
speeds of the two jets are different, one can always find an estimated
distance such that the reciprocal inequality holds, because the system
of equations is under-constrained.

\bibliography{refs}

\begin{thebibliography}{21}
\expandafter\ifx\csname natexlab\endcsname\relax\def\natexlab#1{#1}\fi
\expandafter\ifx\csname bibnamefont\endcsname\relax
  \def\bibnamefont#1{#1}\fi
\expandafter\ifx\csname bibfnamefont\endcsname\relax
  \def\bibfnamefont#1{#1}\fi
\expandafter\ifx\csname citenamefont\endcsname\relax
  \def\citenamefont#1{#1}\fi
\expandafter\ifx\csname url\endcsname\relax
  \def\url#1{\texttt{#1}}\fi
\expandafter\ifx\csname urlprefix\endcsname\relax\def\urlprefix{URL }\fi
\providecommand{\bibinfo}[2]{#2}
\providecommand{\eprint}[2][]{\url{#2}}

\bibitem[{\citenamefont{Ramachandran}(2003)}]{vr}
\bibinfo{author}{\bibfnamefont{V.~S.} \bibnamefont{Ramachandran}},
  \emph{\bibinfo{title}{{The Emerging Mind: Reith Lectures on Neuroscience}}}
  (\bibinfo{publisher}{BBC}, \bibinfo{year}{2003}).

\bibitem[{\citenamefont{Chen et~al.}(2003)\citenamefont{Chen, Friedman, and
  Roe1}}]{mapping}
\bibinfo{author}{\bibfnamefont{L.~M.} \bibnamefont{Chen}},
  \bibinfo{author}{\bibfnamefont{R.~M.} \bibnamefont{Friedman}},
  \bibnamefont{and} \bibinfo{author}{\bibfnamefont{A.~W.} \bibnamefont{Roe1}},
  \bibinfo{journal}{Science} \textbf{\bibinfo{volume}{302}},
  \bibinfo{pages}{881} (\bibinfo{year}{2003}).

\bibitem[{\citenamefont{{Biretta} et~al.}(1999)\citenamefont{{Biretta},
  {Sparks}, and {Macchetto}}}]{M87}
\bibinfo{author}{\bibfnamefont{J.~A.} \bibnamefont{{Biretta}}},
  \bibinfo{author}{\bibfnamefont{W.~B.} \bibnamefont{{Sparks}}},
  \bibnamefont{and}
  \bibinfo{author}{\bibfnamefont{F.}~\bibnamefont{{Macchetto}}},
  \bibinfo{journal}{ApJ} \textbf{\bibinfo{volume}{520}}, \bibinfo{pages}{621}
  (\bibinfo{year}{1999}).

\bibitem[{\citenamefont{{Zensus}}(1997)}]{superluminal2}
\bibinfo{author}{\bibfnamefont{A.~J.} \bibnamefont{{Zensus}}},
  \bibinfo{journal}{{ARA\&A}} \textbf{\bibinfo{volume}{35}},
  \bibinfo{pages}{607} (\bibinfo{year}{1997}).

\bibitem[{\citenamefont{{Einstein}}(1905)}]{einstein}
\bibinfo{author}{\bibfnamefont{A.}~\bibnamefont{{Einstein}}},
  \bibinfo{journal}{Annalen der Physik} \textbf{\bibinfo{volume}{17}},
  \bibinfo{pages}{891} (\bibinfo{year}{1905}).

\bibitem[{\citenamefont{{Mirabel} and {Rodr\'iguez}}(1994)}]{superluminal1}
\bibinfo{author}{\bibfnamefont{I.~F.} \bibnamefont{{Mirabel}}}
  \bibnamefont{and} \bibinfo{author}{\bibfnamefont{L.~F.}
  \bibnamefont{{Rodr\'iguez}}}, \bibinfo{journal}{Nature}
  \textbf{\bibinfo{volume}{371}}, \bibinfo{pages}{46} (\bibinfo{year}{1994}).

\bibitem[{\citenamefont{{Mirabel} and {Rodr\'iguez}}(1999)}]{superluminal3}
\bibinfo{author}{\bibfnamefont{I.~F.} \bibnamefont{{Mirabel}}}
  \bibnamefont{and} \bibinfo{author}{\bibfnamefont{L.~F.}
  \bibnamefont{{Rodr\'iguez}}}, \bibinfo{journal}{{ARA\&A}}
  \textbf{\bibinfo{volume}{37}}, \bibinfo{pages}{409} (\bibinfo{year}{1999}).

\bibitem[{\citenamefont{{Gisler}}(1994)}]{superluminal4}
\bibinfo{author}{\bibfnamefont{G.}~\bibnamefont{{Gisler}}},
  \bibinfo{journal}{Nature} \textbf{\bibinfo{volume}{371}}, \bibinfo{pages}{18}
  (\bibinfo{year}{1994}).

\bibitem[{\citenamefont{{Fender} et~al.}(1999)\citenamefont{{Fender},
  {Garrington}, {McKay}, {Muxlow}, {Pooley}, {Spencer}, {Stirling}, and
  {Waltman}}}]{GRS1915}
\bibinfo{author}{\bibfnamefont{R.~P.} \bibnamefont{{Fender}}},
  \bibinfo{author}{\bibfnamefont{S.~T.} \bibnamefont{{Garrington}}},
  \bibinfo{author}{\bibfnamefont{D.~J.} \bibnamefont{{McKay}}},
  \bibinfo{author}{\bibfnamefont{T.~W.~B.} \bibnamefont{{Muxlow}}},
  \bibinfo{author}{\bibfnamefont{G.~G.} \bibnamefont{{Pooley}}},
  \bibinfo{author}{\bibfnamefont{R.~E.} \bibnamefont{{Spencer}}},
  \bibinfo{author}{\bibfnamefont{A.~M.} \bibnamefont{{Stirling}}},
  \bibnamefont{and} \bibinfo{author}{\bibfnamefont{E.~B.}
  \bibnamefont{{Waltman}}}, \bibinfo{journal}{MNRAS}
  \textbf{\bibinfo{volume}{304}}, \bibinfo{pages}{865} (\bibinfo{year}{1999}).

\bibitem[{\citenamefont{{Rees}}(1966)}]{rees}
\bibinfo{author}{\bibfnamefont{M.}~\bibnamefont{{Rees}}},
  \bibinfo{journal}{{Nature}} \textbf{\bibinfo{volume}{211}},
  \bibinfo{pages}{468} (\bibinfo{year}{1966}).

\bibitem[{\citenamefont{{Perley} et~al.}(1984)\citenamefont{{Perley}, {Dreher},
  and {Cowan}}}]{cyga}
\bibinfo{author}{\bibfnamefont{R.~A.} \bibnamefont{{Perley}}},
  \bibinfo{author}{\bibfnamefont{J.~W.} \bibnamefont{{Dreher}}},
  \bibnamefont{and} \bibinfo{author}{\bibfnamefont{J.~J.}
  \bibnamefont{{Cowan}}}, \bibinfo{journal}{ApJ}
  \textbf{\bibinfo{volume}{285}}, \bibinfo{pages}{L35} (\bibinfo{year}{1984}).

\bibitem[{\citenamefont{{Owsianik} and {Conway}}(1998)}]{hotspot1}
\bibinfo{author}{\bibfnamefont{I.}~\bibnamefont{{Owsianik}}} \bibnamefont{and}
  \bibinfo{author}{\bibfnamefont{J.~E.} \bibnamefont{{Conway}}},
  \bibinfo{journal}{A\&A} \textbf{\bibinfo{volume}{337}}, \bibinfo{pages}{69}
  (\bibinfo{year}{1998}).

\bibitem[{\citenamefont{{Polatidis} et~al.}(2002)\citenamefont{{Polatidis},
  {Conway}, and {Owsianik}}}]{hotspot2}
\bibinfo{author}{\bibfnamefont{A.~G.} \bibnamefont{{Polatidis}}},
  \bibinfo{author}{\bibfnamefont{J.~E.} \bibnamefont{{Conway}}},
  \bibnamefont{and}
  \bibinfo{author}{\bibfnamefont{I.}~\bibnamefont{{Owsianik}}}, in
  \emph{\bibinfo{booktitle}{Proceedings of the 6th European VLBI Network
  Symposium}}, edited by \bibinfo{editor}{\bibnamefont{{{Ros}, {Porcas},
  {Lobanov}, {Zensus}}}} (\bibinfo{year}{2002}).

\bibitem[{\citenamefont{{Jester} et~al.}(2005)\citenamefont{{Jester}, {Roeser},
  {Meisenheimer}, and {Perley}}}]{RF2UV}
\bibinfo{author}{\bibfnamefont{S.}~\bibnamefont{{Jester}}},
  \bibinfo{author}{\bibfnamefont{H.~J.} \bibnamefont{{Roeser}}},
  \bibinfo{author}{\bibfnamefont{K.}~\bibnamefont{{Meisenheimer}}},
  \bibnamefont{and} \bibinfo{author}{\bibfnamefont{R.}~\bibnamefont{{Perley}}},
  \bibinfo{journal}{A\&A} \textbf{\bibinfo{volume}{431}}, \bibinfo{pages}{477}
  (\bibinfo{year}{2005}), \eprint{astro-ph/0410520}.

\bibitem[{\citenamefont{{Piran}}(2002)}]{GRB1}
\bibinfo{author}{\bibfnamefont{T.}~\bibnamefont{{Piran}}},
  \bibinfo{journal}{International Journal of Modern Physics A}
  \textbf{\bibinfo{volume}{17}}, \bibinfo{pages}{2727} (\bibinfo{year}{2002}).

\bibitem[{\citenamefont{Mazets et~al.}(1982)\citenamefont{Mazets, Golenetskii,
  Ilyinskii, Guryan, and Aptekar}}]{GRB5}
\bibinfo{author}{\bibfnamefont{E.~P.} \bibnamefont{Mazets}},
  \bibinfo{author}{\bibfnamefont{S.~V.} \bibnamefont{Golenetskii}},
  \bibinfo{author}{\bibfnamefont{V.~N.} \bibnamefont{Ilyinskii}},
  \bibinfo{author}{\bibfnamefont{Y.~A.} \bibnamefont{Guryan}},
  \bibnamefont{and} \bibinfo{author}{\bibfnamefont{R.~L.}
  \bibnamefont{Aptekar}}, \bibinfo{journal}{Ap\&SS}
  \textbf{\bibinfo{volume}{82}}, \bibinfo{pages}{261} (\bibinfo{year}{1982}).

\bibitem[{\citenamefont{{Piran}}(1999)}]{piran2}
\bibinfo{author}{\bibfnamefont{T.}~\bibnamefont{{Piran}}},
  \bibinfo{journal}{Phys.Rept.} \textbf{\bibinfo{volume}{314}},
  \bibinfo{pages}{575} (\bibinfo{year}{1999}).

\bibitem[{\citenamefont{Ryde}(2005)}]{GRB3}
\bibinfo{author}{\bibfnamefont{F.}~\bibnamefont{Ryde}},
  \bibinfo{journal}{{ApJ}} \textbf{\bibinfo{volume}{614}}, \bibinfo{pages}{827}
  (\bibinfo{year}{2005}).

\bibitem[{\citenamefont{Ryde et~al.}(2003)\citenamefont{Ryde, , and
  Svensson}}]{GRB4}
\bibinfo{author}{\bibfnamefont{F.}~\bibnamefont{Ryde}}, , \bibnamefont{and}
  \bibinfo{author}{\bibfnamefont{R.}~\bibnamefont{Svensson}},
  \bibinfo{journal}{ApJ} \textbf{\bibinfo{volume}{566}}, \bibinfo{pages}{210}
  (\bibinfo{year}{2003}).

\bibitem[{\citenamefont{Ghisellini}(2004)}]{GRB7}
\bibinfo{author}{\bibfnamefont{G.}~\bibnamefont{Ghisellini}},
  \bibinfo{journal}{{J.Mod.Phys.A (Proceedings of 19th European Cosmic Ray
  Symposium - ECRS 2004)}}  (\bibinfo{year}{2004}), \eprint{astro-ph/0411106}.

\bibitem[{\citenamefont{Ryde and Svensson}(2000)}]{GRB6}
\bibinfo{author}{\bibfnamefont{F.}~\bibnamefont{Ryde}} \bibnamefont{and}
  \bibinfo{author}{\bibfnamefont{R.}~\bibnamefont{Svensson}},
  \bibinfo{journal}{ApJ} \textbf{\bibinfo{volume}{529}}, \bibinfo{pages}{L13}
  (\bibinfo{year}{2000}).

\end{thebibliography}

\end{document}